\documentclass{ectj}

\usepackage{amsfonts,amssymb,graphics,epsfig,verbatim,bm,latexsym,amsmath,url,amsbsy}
\usepackage{mathrsfs}
\usepackage{cleveref}
\usepackage{booktabs}
\usepackage{enumitem}
\usepackage{mwe}
\usepackage{pdfpages}
\usepackage{upgreek}

\newtheorem{theorem}{Theorem}
\newtheorem{assumption}{Assumption}
\newtheorem{proposition}{Proposition}

\renewcommand{\thesection}{\arabic{section}}
\renewcommand{\theequation}{\arabic{section}.\arabic{equation}}

\DeclareMathOperator*{\plim}{plim}

\newcounter{bean}
\newcounter{beana}

\received{...}
\accepted{...}
\volume{21}

\title[Feasible IV Regression without Excluded Instruments]{Feasible IV Regression without Excluded Instruments}

\author[E. S. Tsyawo]{Emmanuel Selorm Tsyawo$^{\dagger}$}

\address{$^{\dagger}$AIRESS \& FGSES, Université Mohammed VI Polytechnique, Technopolis-Rabat.  Morocco.}
\email{emmanuel.tsyawo@um6p.ma}

\def\AmSTeX{$\cal A$\kern-.1667em\lower.5ex\hbox{$\cal M$}\kern-.125em
    $\cal S$-\TeX}
\def\BibTeX{{\rm B\kern-.05em{\sc i\kern-.025em b}\kern-.08em
    T\kern-.1667em\lower.7ex\hbox{E}\kern-.125emX}}

\begin{document}

    \begin{abstract}

    The relevance condition of Integrated Conditional Moment (ICM) estimators is significantly weaker than the conventional IV's in at least two respects: (1) consistent estimation without excluded instruments is possible, provided endogenous covariates are non-linearly mean-dependent on exogenous covariates, and (2) endogenous covariates may be uncorrelated with but mean-dependent on instruments. These remarkable properties notwithstanding, multiplicative-kernel ICM estimators suffer diminished identification strength, large bias, and severe size distortions even for a moderately sized instrument vector. This paper proposes a computationally fast linear ICM estimator that better preserves identification strength in the presence of multiple instruments and a test of the ICM relevance condition. Monte Carlo simulations demonstrate a considerably better size control in the presence of multiple instruments and a favourably competitive performance in general. An empirical example illustrates the practical usefulness of the estimator, where estimates remain plausible when no excluded instrument is used.


    \end{abstract}


    \section{Introduction}

Instrumental variable (IV) methods play a pivotal role in the identification and inference on structural parameters in empirical work. When excluded instruments are unavailable or very weak, the usefulness of conventional IV methods is limited. Thanks to the weak relevance condition of Integrated Conditional Moment estimators (ICM hereafter), identification, consistency, and asymptotic normality are attainable even if excluded instruments are unavailable or very weak. These remarkable properties notwithstanding, multiplicative-kernel ICM estimators suffer diminished identification strength with attendant drawbacks, viz. large biases and severe size distortions even for a moderate but fixed number of instruments. In fact, the multiplicative structure of a kernel induces an \emph{artificial weak instrument problem} in the presence of multiple instruments, irrespective of the true strength of non-parametric identification.\footnote{The term ``kernel" in this paper refers to a user-fixed function as in $ U $-statistics -- see, e.g., \citet{lee1990u}.} Moreover, in any given empirical context, the practical usefulness of the aforementioned features of ICM estimators crucially lies in the testability of the ICM relevance condition.

This paper makes a three-fold contribution in light of the foregoing. This paper (1) proposes and investigates the properties of a computationally fast linear ICM estimator which preserves identification strength in the presence of multiple instruments, (2) highlights a remarkable feature of ICM estimators, namely, consistent IV estimation without excluded instruments provided the endogenous covariates are non-linearly mean-dependent on exogenous covariates in a way that ought not to be known or modelled, and (3) extends the test of the relevance condition in \citet{escanciano2018simple} from a single-covariate to a multiple-covariate setting. Although the artificial weak instrument problem is not new in the literature -- e.g., \citet{escanciano2006consistent} -- this paper's characterisation of the problem is, nonetheless, interesting as it provides insight into the case of a moderately large but fixed number of instruments.

The Minimum Mean Dependence estimator (MMD hereafter) proposed in this paper minimises the mean dependence of an error term on a set of instruments using the Martingale Difference Divergence measure (MDD hereafter) of \citet{shao2014martingale}. Thanks to this general notion of dependence, the MMD, like other estimators in the ICM-class, e.g., \citet{dominguez2004consistent}, \citet{escanciano2006consistent}, \citet{shin2008semiparametric}, \citet{antoine2014conditional}, \citet{escanciano2018simple}, \citet{wang2018consistent}, \citet{choi2021generalized}, and \citet{antoine2022partially}, weakens the IV-relevance condition from correlatedness to mean dependence and thus expands the set of relevant IVs in at least two significant ways. First, consistent IV estimation is feasible when there are no excluded instruments, provided endogenous covariates are non-linearly mean-dependent on exogenous covariates in a way that ought not to be known or modelled. This is a valid and non-spurious source of identifying variation as the structural economic model is generally agnostic about the dependence between endogenous and exogenous covariates. Second, as observed in, e.g., \citet{antoine2014conditional}, \citet{escanciano2018simple}, and \citet{antoine2022partially}, endogenous covariates may be weakly correlated or even uncorrelated with but mean-dependent on instruments.

The above features of the ICM-class are not only potentially useful; they are also present in empirical settings. A specification without excluded instruments in the empirical example of this paper yields reasonable estimates and smaller standard errors relative to the IV. The p-value of ICM-relevance, $ 0.08 $, suggests a rejection of the null hypothesis of lack of non-parametric identification at the $ 10\% $ level, although no excluded instrument is used. In Table 1 of \citet{escanciano2018simple}, the standard first-stage $ F $-statistics corresponding to the UK and the US are $ 2.52 $ and $ 2.93 $, respectively, which are indicative of very weak IVs judging by the \citet{stock2005testing} rule-of-thumb of $ 10 $. The respective ICM-relevance p-values, 0.07 and 0.00, are indicative of ICM-strong instruments.

The MMD is identified, $ \sqrt{n} $-consistent, and asymptotically normal. Extensive Monte Carlo simulations in this paper suggest a favourably competitive performance of the MMD within and without the ICM class. This paper shows that the relevance condition of ICM estimators can be cast as an ICM specification test of, e.g.,  \citet{bierens1982consistent}, \citet{bierens1990consistent}, \citet{stute1997nonparametric}, \citet{escanciano2006consistent}, \citet{dominguez2015simple}, \citet{su2017martingale}, and \citet{jiangTsyawo2022aconsistent}. The relevance test in this paper allows a multiple-covariate setting with one endogenous covariate; it thus generalises the single-covariate case of  \citet{escanciano2018simple} -- see Section 4.1 of the paper. A set of simulations in this paper with up to 32 relevant instruments underscore the drawbacks of multiplicative ICM kernels. With sample sizes up to 1000, multiplicative-kernel ICM estimators produce empirical sizes that appear to  increase with the sample size. In fact, three out of four multiplicative-kernel ICM estimators have empirical sizes equal to 1 for a 5\% level $ t $-test at sample sizes 500 and 1000. In contrast, the MMD and the ICM estimator of \citet{escanciano2006consistent}, whose kernels are non-multiplicative and less sensitive to the dimension of the instrument vector, control size meaningfully and improve in performance with the sample size. The computational cost of the  \citet{escanciano2006consistent} estimator relative to the MMD, however, increases exponentially with the sample size.

The MMD complements the class of ICM estimators. ICM estimators extend a literature that is largely focussed on specification testing to estimation. The minimum distance estimator of \citet{wang2018consistent} is the closest to the MMD as both are based on the same kernel. The focus differs, however, as unlike \citet{wang2018consistent} which is concerned about the inconsistency problem highlighted in \citet{dominguez2004consistent}, the current paper specialises to the linear model, provides a closed-form expression of the estimator, and investigates its theoretical properties in detail. Moreover, this paper, unlike \citet{wang2018consistent}, highlights the adverse effects of multiplicative kernels in ICM estimators in the presence of a  moderately sized instrument set. The current paper diverges from the aforementioned in at least three respects. First, available ICM estimators, save \citet{wang2018consistent} and \citet{escanciano2006consistent}, are based on multiplicative kernels. Multiplicative kernels are almost constant at zero and vary negligibly across observations for even moderate dimensions of the instrument -- see \citet[p. 22]{escanciano2022debiased} for a related discussion.\footnote{This paper considers estimation with a finite set of covariates and instruments. Allowing the number of covariates, instruments, or both to grow with the sample size is beyond the scope of the current paper -- see \citet{antoine2022partially}, \citet{escanciano2022debiased}, and \citet{sun2022estimation} for a discussion.} The MMD and \citet{escanciano2006consistent} kernels, in contrast, retain meaningful variation even for a moderately large but finite number of instruments, as shown in the Monte Carlo simulations. The \citet{escanciano2006consistent} estimator is, however, a factor $ O(n) $ more computationally costly than the MMD. A drawback of the MMD is that it requires moment bounds on the set of instruments, unlike bounded-kernel ICM estimators, e.g., \citet{dominguez2004consistent}, \citet{escanciano2006consistent}, \citet{antoine2014conditional}, \citet{escanciano2018simple}, and \citet{choi2021generalized}. Second, the possibility of IV regression without excludability using ICM estimators does not appear to have been noticed before in the literature.\footnote{This generic claim is completely agnostic about the ``first stage" and naturally holds for multiple endogenous covariates without excludability, unlike the 2017 working paper version of \citet{choi2021generalized}.} Third, this paper provides a generally useful test of the ICM non-parametric identification condition in a multiple-covariate setting.

The rest of the paper is organised as follows. \Cref{Sect:Measure_Estimator} presents the MMD estimator, and  \Cref{Sect:Asymptotic_Theory} collects theoretical results viz. consistency, asymptotic normality, consistency of the covariance matrix estimator, and the ICM relevance test. \Cref{Sect:Monte_Carlo_Sim} examines the small sample performance of the estimator and its covariance matrix estimator via simulations. An empirical example in \Cref{Sect:Empirical_Example} illustrates the proposed estimator's practical usefulness, and \Cref{Sect:Conclusion} concludes. All proofs and additional simulation results are relegated to the appendix and the online supplement.

\section{The MMD Estimator}\label{Sect:Measure_Estimator}

\subsection{The MDD measure}
\setcounter{equation}{0}

Consider the econometric model of interest: $ Y = X\theta_o + U $ where $ Y $ is the outcome, elements in the vector of covariates $ X \in \mathbb{R}^{p_x} $ may be endogenous, $ Z \in \mathbb{R}^{p_z} $ collects instrumental variables, $  E[U|Z] = 0 \ a.s. $, and $ \theta_o $ is the parameter vector of interest. $ X $ contains a constant term which ensures the error term $ U $ is zero-mean. The conventional IV uses the orthogonality condition $  E[UZ] = 0 $ which, as observed by \citet{dominguez2004consistent}, fails to fully exploit the stated conditional moment restriction while an ICM estimator directly solves the conditional moment restriction $  E[U|Z] = 0 \ a.s. $ ICM measures convert the  possibly continuum of orthogonality conditions implied by $  E[U|Z] = 0 \ a.s. $ to a scalar-valued objective function which is zero if and only if $  E[U|Z] = 0 \ a.s. $ (\citealp{bierens1982consistent}). As the proposed MMD is based on the Martingale Difference Divergence (MDD) ICM measure of \citet{shao2014martingale}, it is instructive to briefly present the measure. 

For a zero-mean $ W \in \mathbb{R} $ and a multivariate random variable $ Z \in \mathbb{R}^{p_z} $, the MDD is defined as the square root of
\[
\mathrm{MDD}^2(W|Z) \equiv \frac{1}{c_{p_z}} \int \frac{|E[W\exp(\iota Zs)]|^2}{||s||^{1+p_z}}ds
\]
where $ c_{p_z} = \frac{\pi^{(1 + p_z)/2}}{\Gamma((1 + p_z)/2)} $, $ \Gamma(\cdot) $ is the complete gamma function $ \Gamma(p) \equiv \int_{0}^{\infty} t^{p-1}\exp(-t)dt  $, $ \iota = \sqrt{-1} $ is the imaginary unit, and $ ||\cdot|| $ denotes the Euclidean norm. The MDD uses the non-integrable weight function $ w(s) = 1/(c_{p_z}||s||^{1+p_z}) $ proposed by \citet{szekely2007measuring}.

The MDD belongs to the general class of ICM measures of mean dependence with the form 
\begin{equation}\label{eqn:MDD_general}
	\mathrm{GMDD}^2(W|Z) \equiv \int| E[WG(Z,s)]|^2w(s)ds
\end{equation}
\noindent where different suitable choices of $ G(Z,s) $  and $ w(s) $ give rise to different ICM measures. For example, \citet{escanciano2018simple} uses the complex exponential $ G(Z,s) = \exp(\iota Zs) $ with the standard normal probability density function as weight $ w(s) $  -- see \citet{dominguez2004consistent}, \citet{escanciano2006consistent}, \citet{bierens1990consistent}, and \citet{kim2020robust} for other examples. This paper's choice of the pair $ G(Z,s) = \exp(\iota Zs) $ and $ w(s) = 1/(c_{p_z}||s||^{1+p_z}) $ is borne out of a natural double requirement: tractability (i.e., without requiring numerical integration) and informativeness of the resulting ICM measure even for moderately large $ p_z $ -- the latter is discussed in \Cref{Sub_Sect:ACK_Problem}.

For ease of reference, the following theoretical results on the MDD measure are collected below as properties; see \citet{shao2014martingale} for details.
	\setcounter{bean}{0}
	\begin{list}
		{(\alph{beana})}{\usecounter{beana}}
		\item $ \mathrm{MDD}(W|Z) \geq 0 $;
		\item $ \mathrm{MDD}(W|Z) = 0 $ if and only if $ W $ is mean-independent of $ Z $ under the condition $  E[||W||^2+||Z||^2] < \infty $; and
		\item $ \mathrm{MDD}^2(W|Z) = - E[||Z - Z^\dagger||WW^\dagger] $ under the condition $  E[||W||^2+||Z||^2] < \infty $ where $ [W^\dagger,Z^\dagger] $ is an independent and identically distributed ($ iid $) copy of $ [W,Z] $.
	\end{list}

\subsection{Computation}
The analytical expression of the MDD measure in Property (c) is preferred in setting up the objective function as it avoids working with the integral or derivative of the norm of a complex number. The expected value of the objective function is thus 
\begin{equation}\label{eqn:Obj_Fun}
	Q_o(\theta) = - E[||Z-Z^\dagger||(Y - X\theta)(Y^\dagger - X^\dagger\theta)]
\end{equation} where $ [Y^\dagger,X^\dagger,Z^\dagger] $ is an $ iid $ copy of $ [Y,X,Z] $. For a random variable indexed by $ i $ or $ (i,j) $, define $  E_n[\xi_i] \equiv \frac{1}{n}\sum_{i=1}^{n}\xi_i $ and $  E_n[\xi_{ij}] \equiv \frac{1}{n^2}\sum_{i=1}^{n}\sum_{j=1}^{n}\xi_{ij} $. The sample version of $ Q_o(\theta) $ is given by 
\[
Q_n(\theta) = - E_n[||Z_i - Z_j||(Y_i - X_i\theta)(Y_j - X_j\theta)].
\]
The following result shows that the MMD is a linear IV estimator. 

\begin{proposition}\label{Prop:IV_Estimator}
	Let $ \theta_o $ be the minimiser of $ Q_o(\theta) $ in \eqref{eqn:Obj_Fun}, then $ \hat{\theta}_n $, its estimator, is a linear IV estimator given by the closed-form expression \begin{equation}\label{eqn:Estimator}
		\hat{\theta}_n = \big( E_n[h_n(Z_i)'X_i]\big)^{-1}  E_n[h_n(Z_i)'Y_i]
	\end{equation} where $ h_n(Z_i) \equiv \frac{1}{n-1}\sum_{j= 1}^{n}||Z_i-Z_j||X_j $ is a $ 1 \times p_x $ row-vector of (constructed) instruments.
\end{proposition}

\noindent $ \hat{\theta}_n $ in \eqref{eqn:Estimator} is a linear IV-estimator -- see, e.g., \citet[p. 92]{wooldridge2010econometric} -- with the constructed instrument $ \{h_n(Z_i), 1 \leq i \leq n\} $. It remains an IV estimator irrespective of the dimension of $ Z $ since $ ||Z_i - Z_j|| $ remains scalar-valued in the ``over-identified" ($ p_z > p_x $), ``just-identified" ($ p_z=p_x $), or ``under-identified" ($ p_z < p_x $) case. Estimation thus  proceeds by constructing $ \{h_n(Z_i), 1 \leq i \leq n\} $ and running a standard IV regression. It is shown in \Cref{Sect:Asymptotic_Theory} that inference follows through exactly as the standard IV. Unlike the ICM estimators of \citet{antoine2014conditional}, \citet{escanciano2018simple}, and  \citet{choi2021generalized}, where scale invariance is induced by scaling $ Z $, the MMD is not only naturally scale invariant but also invariant to affine transformations after rotation.\footnote{This is referred to as rigid motion invariance in the Statistics literature -- see, e.g., \citet[Sect. 2]{szekely2012uniqueness} and \citet[Theorem 1.4]{shao2014martingale}.} For all $ c, q  \in \mathbb{R} $, $q\not = 0 $, orthonormal $ Q \in \mathbb{R}^{p_z\times p_z} $, and affine transformed $ Z $ after a rotation, i.e., $ \tilde{Z} \equiv c+qZQ $, $ \mathrm{MDD}^2(U|\tilde{Z}) = - E[||qZQ - qZ^\dagger Q||UU^\dagger] = |q|\mathrm{MDD}^2(U|Z) $ since $ ||ZQ-Z^\dagger Q|| = \sqrt{(Z-Z^\dagger )QQ'(Z-Z^\dagger )'}=||Z-Z^\dagger|| $.

\subsection{The MMD in the ICM-class of estimators}
The MMD belongs to the class of linear ICM estimators. Estimators in this class can be cast in the form \eqref{eqn:Estimator}; they differ by the kernel $ K(Z_i,Z_j) $ which replaces $ ||Z_i-Z_j|| $ in \eqref{eqn:Estimator}. Kernel functions result from the specific combination of $ G(Z,s) $ and $ w(s) $ in \eqref{eqn:MDD_general}. To see this, note that the generalised MDD measure \eqref{eqn:MDD_general} is 
\[
\int| E[WG(Z,s)]|^2w(s)ds =  E[WW^\dagger \int G(Z,s)\overline{G(Z^\dagger,s)}w(s)ds] =  E[WW^\dagger K(Z,Z^\dagger)]
\]
where $ \overline{\xi} $ denotes the complex conjugate of $ \xi $.\footnote{$ \overline{G(Z^\dagger,s)} = G(Z^\dagger,s) $ for real-valued $ G(Z^\dagger,s) $.} Examples of kernels used in ICM estimators include $ K(Z_i,Z_j) = \exp(-0.5(Z_i-Z_j)\hat{V}_z^{-1}(Z_i-Z_j)')$ in the Integrated Instrumental Variable (IIV) estimator of \citet{escanciano2018simple} where $ \hat{V}_z $ is the sample covariance of $ \{Z_i, 1\leq i \leq n\} $, $ K_n(Z_i,Z_j) = \frac{1}{n}\sum_{l=1}^{n} \mathrm{I}(Z_i \leq Z_l )\mathrm{I}(Z_j \leq Z_l) $ in \citet{dominguez2004consistent}, and  
\[
K_n(Z_i,Z_j) = \frac{\pi^{(p_z/2)-1}}{\Gamma((p_z/2)+1)}\sum_{l=1}^{n} A_{ijl}^{(0)}/n \text{ where } A_{ijl}^{(0)} = \Big| \pi - \mathrm{arcos}\Big(\frac{(Z_i-Z_l)\cdot(Z_j-Z_l)}{||Z_i-Z_l||\times||Z_j-Z_l||}\Big) \Big|,
\] and $ Z\cdot Z^\dagger $ denotes the dot product of $ Z $ and $ Z^\dagger $ in \citet{escanciano2006consistent}.\footnote{The subscript $ n $ on the \citet{dominguez2004consistent} and \citet{escanciano2006consistent} kernels is meant to emphasise dependence on the sample.} On the estimator of \citet{escanciano2006consistent}, the following computational simplifications apply: $ A_{ijl}^{(0)} = \pi $ if $ Z_l\not=Z_i = Z_j $, $ Z_l=Z_i\not=Z_j $, or $ Z_l=Z_j\not=Z_i $, and $ A_{ijl}^{(0)} = 2\pi $ if $ Z_l = Z_i = Z_j $ -- \citet[Appendix B]{escanciano2006consistent}. Let $ Y_i^* \equiv [Y_i, X_i] $ and $ \tilde{K}(Z_i-Z_j) $ be a symmetric and bounded density if $ i\not =j $ and zero otherwise. $ K(Z_i,Z_j) $ equals $ \tilde{K}(Z_i-Z_j) $ if $ i\not=j $ and $ -\tilde{\lambda} $ otherwise in the Weighted Minimum Distance (WMD) of \citet{antoine2014conditional}, where $ \tilde{\lambda} $ equals the smallest eigen-value of $ ( E_n[{Y_i^*}'Y_i^*])^{-1} E_n[\tilde{K}(Z_i-Z_j){Y_i^*}'Y_j^*] $. Although fundamentally ICM, the \citet{antoine2014conditional} kernel has an induced jackknifed limited information maximum likelihood (LIML)-like structure. Although the kernels of \citet{dominguez2004consistent} and  \citet{escanciano2006consistent} (DL and ESC6 respectively hereafter) involve quite natural choices of weight $ w(s) $ such as the uniform density on the unit sphere, the empirical distribution of the data, or both, they come at an $ O(n^3) $ computational cost relative to the MMD or the IIV's $ O(n^2) $ for example. The aforementioned kernels, save the MMD's $ K(Z_i,Z_j) = -||Z_i-Z_j|| $ and ESC6's, are bounded multiplicative kernels. It is easily verified that the ESC6 kernel is also invariant to affine transformations after rotation.

\subsection{The almost constant kernel problem}\label{Sub_Sect:ACK_Problem}

As instruments $ Z $ enter the ICM measure \eqref{eqn:MDD_general} through the kernel only, the variation of the kernel across varying values of $ Z $ is crucial for the measure to be informative of mean dependence. A useful characterisation of the dependence of the kernel on the dimension of $ Z $ can be achieved through appropriate means $ \zeta $, e.g., quadratic, geometric, or arithmetic, of measurable functions of $ [Z,Z^\dagger] $ as presented in \Cref{Tab:Kernel_Characterise}.\footnote{See the online supplement for details.} The characterisation is useful as it highlights the sensitivity of the kernel to $ p_z $.
\begin{table}[!htbp]
	\centering
	\caption{Characterisation of Kernels}
	\label{Tab:Kernel_Characterise}
	\begin{tabular}{@{}lllcc@{}}
		\hline\hline
		\multicolumn{2}{l}{Kernel} & Characterisation & $\zeta$ & Support of $\zeta$ \\ \hline
		MMD            &  & $ K(Z,Z^\dagger)= -\zeta\sqrt{p_z} $ & $ \sqrt{\frac{1}{p_z} \sum_{k=1}^{p_z} (Z_k-Z_k^\dagger)^2 } $ & $ [0,\infty) $ \\
		Gauss           &  & $ K(Z,Z^\dagger) = \zeta^{p_z} $ & $ \Big(\prod_{k=1}^{p_z} \exp(-0.5(Z_k-Z_k^\dagger)^2)\Big)^{1/p_z} $ & $ [0,1] $ \\
		ESC6           &  & $ K(Z,Z^\dagger) \leq \zeta/\sqrt{p_z} $ & $\displaystyle \Big(\frac{\sqrt{\pi}\mathrm{e}}{3}\Big)^3\lim_{n\rightarrow\infty}\sum_{l=1}^{n} A_{ijl}^{(0)}/n $ & $ \big[0, 2\pi \big(\frac{\sqrt{\pi}\mathrm{e}}{3}\big)^3\big] $ \\ \hline\hline
	\end{tabular}
\end{table}The \citet{antoine2014conditional} and \citet{escanciano2018simple} kernels are not separately presented in \Cref{Tab:Kernel_Characterise} as they are modifications of the Gaussian kernel $ K(Z,Z^\dagger)=\exp(-0.5||Z-Z^\dagger||^2) $. The almost sure limit of the DL kernel suggests that its characterisation depends on the joint distribution of $ Z $.\footnote{$ K_n(Z,Z^\dagger) \xrightarrow{a.s.} 1 - F_Z(Z\vee Z^\dagger)  $ for the DL kernel where $ F_Z(\cdot) $ denotes the joint cumulative distribution function of $ Z $ and $ Z\vee Z^\dagger $ denotes the element-wise maxima of the vectors $ Z $ and $ Z^\dagger $. For example, the DL kernel with a multivariate normal $ Z $ has the Gaussian characterisation in \Cref{Tab:Kernel_Characterise} with element-wise maxima $ Z\vee Z^\dagger $ as $ n\rightarrow \infty $ almost surely (a.s.). That of the ESC6 is expressed using the angular distance in \citet{kim2020robust}.} In general, kernels with the multiplicative structure have the characterisation $ K(Z,Z^\dagger) \equiv \prod_{k=1}^{p_z} \varphi(Z_k,Z_k^\dagger) = \zeta_{m}^{p_z}, $ where $ \zeta_{m} \equiv \Big(\prod_{k=1}^{p_z} \varphi(Z_k,Z_k^\dagger)\Big)^{1/p_z}  $ is the geometric mean of $ \{ \varphi(Z_k,Z_k^\dagger), \ 1\leq k \leq p_z \} $ and $ \varphi(Z_k,Z_k^\dagger) $ is a bounded function, e.g., the normal probability density function which gives a Gaussian kernel. When $ \varphi(Z_k,Z_k^\dagger) \in [0,1] $ for each $ k \in \{1,\ldots, p_z\} $ in particular, $ \mathrm{var}(K(Z,Z^\dagger)) \leq  E[\prod_{k=1}^{p_z} \varphi(Z_k,Z_k^\dagger)^2] =  E[\zeta_{m}^{2p_z}] $ can be very small even for moderate $ p_z $, say, $ p_z=18 $. Multiplicative kernels $ K(Z,Z^\dagger) $ tend to have negligible variation and thus result in an \emph{almost constant kernel problem} even for a moderate dimension of $ Z $ as they are very sensitive to the dimension of $ Z $. The structure of the MMD and ESC6 kernels, in contrast, allows for an informative measure as both kernels are less sensitive to $ p_z $. To shed further light on the problem, note that by the Cauchy-Schwartz inequality, an ICM measure $ \mathrm{GMDD}^2(W|Z) $ satisfies 
\begin{equation}\label{eqn:GMDD_Cov_CS}
0 \leq \mathrm{GMDD}^2(W|Z) = \mathrm{cov}[K(Z,Z^\dagger),WW^\dagger]\leq  E[W^2] \sqrt{\mathrm{var}[K(Z,Z^\dagger)]}.
\end{equation} Irrespective of the strength of mean dependence, $ \mathrm{GMDD}^2(W|Z) $ can be approximately zero even for a moderate $ p_z $ as the variation of $ K(Z,Z^\dagger) $ is approximately zero. As the GMDD is a covariance (see \eqref{eqn:GMDD_Cov_CS}) and the scale of $ K(Z,Z^\dagger) $ varies by kernel and instruments $ Z $, a correlation version of the GMDD (see, e.g., \citet[p. 1304]{shao2014martingale}) is unit- and scale-free and ensures comparability across different measures' informativeness of mean dependence.

\begin{figure}
	\centering
	\begin{minipage}{0.45\textwidth}
		\centering
		\includegraphics[width=0.9\textwidth]{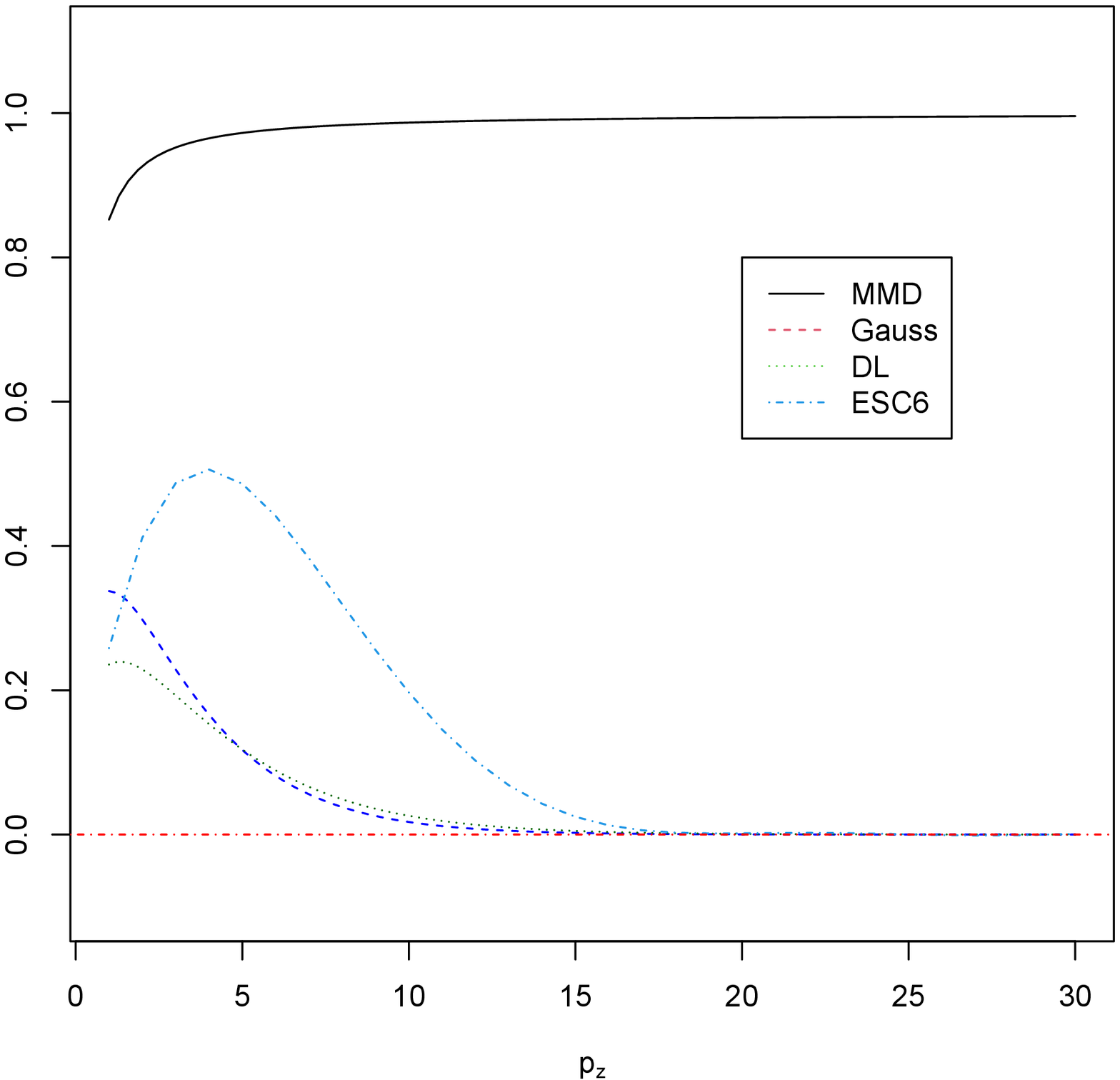} 
		\caption{Standard Deviation of Kernels}
		\label{Fig:Kernel_Var}
	\end{minipage}\hfill
	\begin{minipage}{0.45\textwidth}
		\centering
		\includegraphics[width=0.9\textwidth]{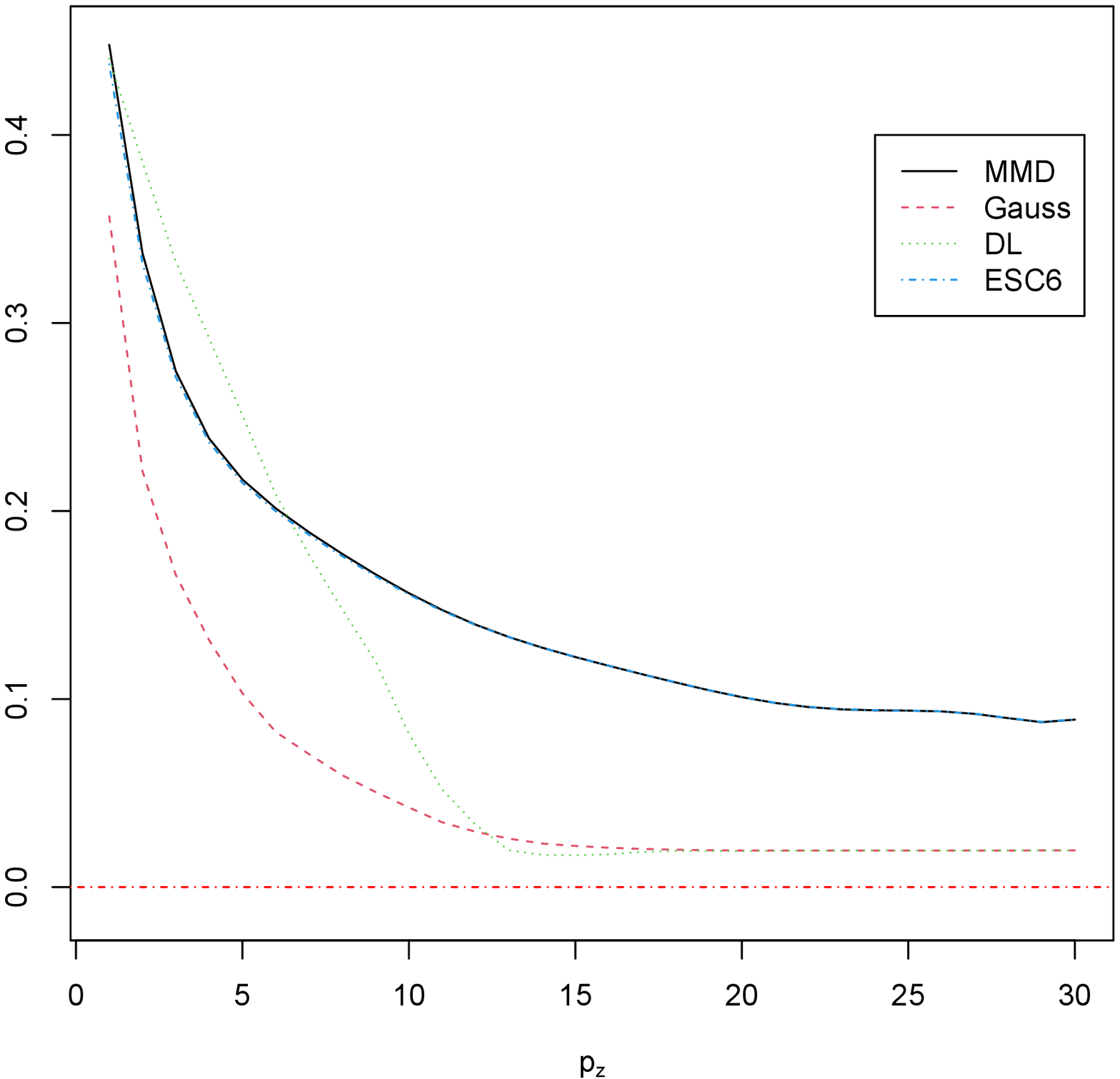} 
		\caption{GMDC of Kernels}
		\label{Fig:GMDC}
	\end{minipage}
\end{figure}

Consider an illustrative example with $ Z \sim \mathcal{N}(0,I_{p_z}) $. The plot of the standard deviations of the kernels in \Cref{Fig:Kernel_Var} clearly shows that an ICM measure using the Gaussian or DL kernel can be very small, irrespective of the actual strength of mean dependence.\footnote{For $ E[W^2]=1 $ and $ p_z = 18 $, $ \mathrm{GMDD}^2(W|Z) \leq (\mathrm{var}[K(Z,Z^\dagger)])^{1/2} \approx 7.1\times 10^{-4} $ for the Gaussian kernel while $ \mathrm{MDD}^2(W|Z) \leq (\mathrm{var}[K(Z,Z^\dagger)])^{1/2} \approx 0.99 $. See the online supplement for details.} In contrast, a measure based on the MMD kernel enjoys meaningful variation even for moderate $ p_z $. To illustrate the informativeness of the measure of mean dependence, \Cref{Fig:GMDC} plots the Generalised Martingale Difference Correlation (GMDC) -- a correlation analogue of the GMDD -- to ensure comparability of measures' informativeness across different kernels; see Section S2 of the online supplement for details.\footnote{The mean dependence of $ W $ on $ Z $ is modelled as $  E[W|Z] = \frac{1}{\sqrt{p_z}}\sum_{k=1}^{p_z}Z_k $, and the GMDCs are simulated.} One observes that for the same level of mean dependence of $ W $ on $ Z $, GMDCs based on  multiplicative kernels are less informative of the strength of mean dependence relative to the MDD and ESC6. Moreover, notice that GMDCs of both the MDD and ESC6 coincide. This is not surprising as the characterisation of both kernels in \Cref{Tab:Kernel_Characterise} suggests that the ratio of their GMDCs is approximately 1.

\section{Asymptotic Theory}\label{Sect:Asymptotic_Theory}
\setcounter{theorem}{0}
\setcounter{equation}{0}

Regularity conditions for the MMD are grouped into two categories. The first category (Assumptions \ref{Ass:DGP} and \ref{Ass:Bounds}) comprises sampling and dominance conditions.

\begin{assumption}\label{Ass:DGP}
	Observed data $ \{[Y_i,X_i,Z_i], 1 \leq i \leq n\} $ are independently and identically distributed ($ iid $)  with the outcome generated as $ Y_i =  X_i\theta_o + U_i $.
\end{assumption}

\begin{assumption}\label{Ass:Bounds}
	(a) There exists a positive constant $ C < \infty $ such that $  E||X||^4\leq C $ and $  E||Z||^4\leq C $; (b) $ 0 < \sigma^2(Z) \equiv   E[U^2|Z] < \bar{\sigma}^2 $ a.s. for some positive constant $ \bar{\sigma}^2 < \infty$.
\end{assumption}
\noindent The $ iid $ setting in \Cref{Ass:DGP} allows for conditional heteroskedasticity $ \sigma^2(Z) \equiv   E[U^2|Z] $ of unknown form.  The boundedness assumptions are sufficient for the existence of the MDD measure (see \citet[Theorem 1]{shao2014martingale}) and useful in establishing the asymptotic properties of the MMD estimator. 

An unbounded MMD kernel comes at the cost of imposing a bound on the fourth moment of $ Z $ in \Cref{Ass:Bounds}(a). The MMD may therefore be sensitive to outliers in $ Z $. This is, however, no limitation of the MMD as $ Z $ can simply be replaced by an element-wise bounded one-to-one mapping such that $ Z $ and its mapping generate the same Euclidean Borel field, e.g., $ \mathrm{atan}(Z) $ -- see \citet[p. 108]{bierens1982consistent}. The moment bound on $ ||Z|| $ is not necessary for bounded-kernel ICM estimators as the instruments $ Z $ enter the estimator only through the bounded kernel -- see, e.g., \citet[Assumption 4]{antoine2014conditional} or \citet[Sect. 3]{kim2020robust}. This relative advantage of bounded-kernel ICM estimators only applies to excluded instruments in $ Z $ as included instruments in $ X $ ought to obey the moment bounds imposed in \Cref{Ass:Bounds}(a) -- cf. \citet[pp. 37 and 41]{escanciano2018simple}. In effect, this advantage no longer holds when $ Z $ contains no excluded instrument. The foregoing also suggests that the MMD is not sensitive to outliers in $ Y $ or $ X $ any more than bounded-kernel ICM estimators. The bound on $ Z $ in \Cref{Ass:Bounds}(a) is standard in the IV context -- cf.  \citet[Assumption 12.2.3]{hansen_2021econometrics}. \Cref{Ass:Bounds}(b) places a standard condition on the error term $ U $, e.g., \citet[Assumption 3]{hansen2014instrumental} and \citet[Assumption in Proposition 1]{carrasco2015regularized}.

\subsection{Identification}
The second set of regularity conditions is needed for identification; they are conventional IV-type identification conditions which ensure that the expected value of the objective function \eqref{eqn:Obj_Fun} is uniquely minimised at $ \theta_o $. Define $ m(z) \equiv  E[X|Z=z] $.
\begin{assumption}[Identification]\label{Ass:Identif}
	(a) $  E[U|Z] = 0 \ a.s $; (b) $  E[m(Z)'m(Z)] $ is positive definite.
\end{assumption}

\noindent \Cref{Ass:Identif}(a) and \Cref{Ass:Identif}(b) are, respectively, ICM analogues of IV exclusion and relevance identification conditions. 

\Cref{Ass:Identif}(a) is a standard exogeneity condition for ICM estimators, e.g., \citet[Assumption 1(ii)]{antoine2014conditional} and \citet[eqn. 1.2]{escanciano2018simple}. Although stronger than the IV exclusion restriction ($  E[Z'U] = 0 $) when $ X=Z $ or $ Z $ contains excluded instruments, \Cref{Ass:Identif}(a) can be viewed as weak in the sense that $ Z $ may simply include strictly exogenous elements in $ X $ without any excluded instruments provided \Cref{Ass:Identif}(b) holds --  \Cref{Ass:Identif}(b) can hold even for $ p_z<p_x $. \Cref{Ass:Identif}(a) is typically tested using ICM specification tests.

\Cref{Ass:Identif}(b) is an ICM relevance condition -- see, e.g., \citet[Assumption 1.1]{escanciano2018simple}. It can be equivalently expressed as a \emph{linear completeness} condition (LC hereafter) $  E[X|Z]\tau = 0 \implies \tau = 0 $ a.s. as termed by \citet{escanciano2018simple}. \Cref{Ass:Identif}(b) appears as an identification rank condition within the context of the non-parametric IV (\citet[Assumption 2(i)]{donald2001choosing}), as a \emph{Local Identifiability} condition in the context of a non-linear ICM estimator (\citet[Assumption 7]{antoine2014conditional}), and as a relevance condition in the context of a partially linear ICM estimator (\citet[Assumption 2.1(ii)]{antoine2022partially}) -- see  \citet{newey2003instrumental} for a related discussion in the context of the fully non-parametric IV. The LC formulation of \Cref{Ass:Identif}(b) is perhaps a more interesting interpretation of ICM relevance; it says no non-zero linear combination of $ X $ is mean-independent of $ Z $.\footnote{In the simple case of a univariate $ X $, \Cref{Ass:Identif}(b) simply says $ X $ is mean-dependent on $ Z $. This is the case covered by the test in \citet[Sect. 4.1]{escanciano2018simple}.} The LC test introduced in this paper draws on this insight. Although the completeness condition in the fully non-parametric context is not testable (\citealp{canay2013testability}), the LC condition of ICM estimators is testable (\citealp{escanciano2018simple}). Both ICM identification conditions in \Cref{Ass:Identif} are hence testable. An analogous interpretation for the standard IV full rank (relevance) condition on $  E[Z'X] $, see, e.g., \citet[eqn. 5.10]{wooldridge2010econometric}, is that no non-zero linear combination of $ X $ is uncorrelated with $ Z $. Since mean independence implies uncorrelatedness, one intuitively sees why \Cref{Prop:WR_Rank_Weak_Cond}(c) below holds. One may wonder how \Cref{Ass:Identif}(b) translates into the context of the MMD and its specific kernel. This is addressed in \Cref{Prop:WR_Rank_Weak_Cond}(b) -- cf. \citet[Proposition 2.1]{escanciano2018simple}.

Define the function $ h(z) \equiv  E[||z-Z||X] $. The matrix $ - E[||Z-Z^{\dagger}||X'X^{\dagger}] $ can be shown to equal $ - E[h(Z)'X] $, and it is the expected value of $ - E_n[h_n(Z_i)'X_i] = - E_n[||Z_i-Z_j||X_j'X_i] $ in the MMD estimator \eqref{eqn:Estimator}. Moreover, it is real, symmetric, and positive semi-definite. To see why it is positive semi-definite, observe that for all $ \uptau \in \mathcal{S}_{p_x} $ where $ \mathcal{S}_p \equiv \{ \uptau \in \mathbb{R}^p: ||\uptau|| = 1 \} $ denotes the space of vectors $ \uptau \in \mathbb{R}^p $ with unit Euclidean norm, 
\begin{equation}\label{eqn:MDDM_psd}
	-\uptau' E[||Z-Z^{\dagger}||X'X^{\dagger}]\uptau = - E[||Z-Z^{\dagger}||(X\uptau)(X^{\dagger}\uptau)] = \frac{1}{c_{p_z}} \int \frac{| E[X\uptau\exp(\iota Zs)]|^2}{||s||^{1+p_z}}ds \geq 0.
\end{equation}

\begin{proposition}\label{Prop:WR_Rank_Weak_Cond} 
	(a) $ \theta_o $ is identified under \Cref{Ass:Identif}; (b) \Cref{Ass:Identif}(b) and the condition $ p_x = \mathrm{rank}\big( E[h(Z)'X]\big) $ are equivalent; (c) $p_x \geq \mathrm{rank}\big( E[h(Z)'X]\big) \geq \mathrm{rank}\big( E[Z'X]\big) $.
\end{proposition}
From the proof of \Cref{Prop:WR_Rank_Weak_Cond}(a), $ Q_o(\theta) - Q_o(\theta_o) = -  (\theta-\theta_o)' E[h(Z)'X](\theta-\theta_o) $, thus $ - E[h(Z)'X] = - E[||Z-Z^{\dagger}||X'X^{\dagger}] $ informs the strength of identification. Since $  E[X\uptau|Z] = 0 $ is equivalent to $  E[(X_{-1}- E[X_{-1}])\uptau_{-1}|Z] = 0 $ for $ \uptau \in \mathcal{S}_{p_x} $, $ \uptau_{-1} \in \mathcal{S}_{p_x-1} $, and $ X_{-1}\in \mathbb{R}^{p_x-1} $ which is $ X $ with the constant term excluded, the condition $ p_x-1 = \mathrm{rank}\big( E[||Z-Z^{\dagger}||(X_{-1}- E[X_{-1}])'(X_{-1}^{\dagger}- E[X_{-1}])]\big) $ is  equivalent to \Cref{Ass:Identif}(b) in view of \Cref{Prop:WR_Rank_Weak_Cond}(b). From the foregoing, 
\[
\uptau_{-1}' E[K(Z-Z^{\dagger})(X_{-1}- E[X_{-1}])'(X_{-1}^{\dagger}- E[X_{-1}])]\uptau_{-1} =  \mathrm{GMDD}^2(X_{-1}\uptau_{-1}|Z) \geq 0
\]
whence  
\[
0\leq \uptau_{-1}' E[K(Z,Z^{\dagger})||(X_{-1}- E[X_{-1}])'(X_{-1}^{\dagger}- E[X_{-1}])]\uptau_{-1} 
\leq \mathrm{var}[X_{-1}\uptau_{-1}]\sqrt{\mathrm{var}[K(Z,Z^{\dagger})]}
\] 
by \eqref{eqn:GMDD_Cov_CS} for any estimator in the ICM-class with a kernel $ K(Z,Z^\dagger) $. The above suggests the following unit- and scale-free measure of ICM identification strength: $ \mathrm{GMDC}(X_{-1}\uptau_{-1}^*|Z) $ where $ \uptau_{-1}^* $ is the eigen-vector associated with the smallest eigen-value of $  E[K(Z,Z^{\dagger})||(X_{-1}- E[X_{-1}])'(X_{-1}^{\dagger}- E[X_{-1}])] $.  Under a standard moment boundedness condition on $ X $, e.g., \Cref{Ass:Bounds}(a), all eigen-values of $  E[K(Z,Z^{\dagger}) X'X^{\dagger}] $ can be very close to zero if  $ \sqrt{\mathrm{var}[K(Z,Z^{\dagger})]} $ is. The almost constant kernel problem thus hurts identification in the context of ICM estimators since in the presence of multiple instruments, it is possible that $ \mathrm{GMDC}(X\uptau|Z) \approx 0 $ but $  E[X|Z]\uptau \not = 0 $ a.s. for some $ \uptau \in \mathcal{S}_{p_x} $. This implies the equivalence relationship in \Cref{Prop:WR_Rank_Weak_Cond}(b) is weakened for multiplicative-kernel ICM estimators, relative to the MMD and ESC6 for example, even for a moderate $ p_z $. In practice, the almost constant kernel problem induces an \emph{artificial weak instrument} problem in multiplicative-kernel ICM estimators, irrespective of the actual strength of non-parametric ICM identification.

\Cref{Prop:WR_Rank_Weak_Cond}(c) shows that the LC condition \Cref{Ass:Identif}(b) is weaker than the standard IV relevance condition. The weakening can be viewed along two dimensions. First, the dimension of $ Z $ can be less than that of $ X $, i.e., there can be fewer instruments than covariates. This is under-identification in the conventional IV setting.\footnote{This feature can be gleaned from \citet[Example 2]{dominguez2004consistent} with $ X = [Z,Z^2] $ and univariate $ Z $. The authors, however, did not consider the general case in this paper where $ X $ may be an unknown function of $ Z $ or where some covariates in $ X $ are endogenous. \citet{sun2022estimation} also relies on this feature to estimate the parameters on two endogenous covariates using only one excluded instrument.} An important implication of this is that a researcher who faces an endogeneity problem but lacks excluded instruments can still consistently estimate $ \theta_o $ provided the mean of endogenous covariates in $ X $ have some non-linear dependence on the exogenous covariates in $ X $. Take $ X = [D,Z] $, $ D = Z + Z^2 + V $, $  E[V|Z]=0 $, and $ Z \sim \mathcal{N}(0,1) $ as an illustrative example. There is no non-zero linear combination such that $  E[X\uptau|Z] = (\uptau_1+\uptau_2)Z + \uptau_1Z^2$ is not a function of $ Z $. \Cref{Ass:Identif}(b) is  thus satisfied without an excluded instrument. Second, endogenous covariates in $ X $ can be weakly correlated or uncorrelated with but mean-dependent on $ Z $ as observed in, e.g., \citet{antoine2014conditional}, \citet{escanciano2018simple}, and \citet{antoine2022partially}.

\subsection{Consistency, Asymptotic Normality, and Covariance Matrix Estimation}
Define $ A \equiv - E[h(Z)'X] $, $ \hat{A}_n \equiv  -E_n [h_n(Z_i)'X_i] $, and $ B \equiv \mathrm{var}\big[\sqrt{n} E_n[h(Z_i)'U_i] \big] $. It is shown in the online supplement that $ \hat{A}_n \xrightarrow{a.s.} A $. Combining the expression of the MMD estimator \eqref{eqn:Estimator}, the data generating process (\Cref{Ass:DGP}), and the continuous mapping theorem, $ \hat{\theta}_n $ satisfies the following expansion: 
\begin{equation}\label{eqn:Estimator_Expansion}
	\sqrt{n}(\hat{\theta}_n - \theta_o)  = (A^{-1} + o_p(1))\sqrt{n} E_n[h_n(Z_i)'U_i].
\end{equation}
\noindent It is shown in the online supplement that $  E_n[(h_n(Z_i) - h(Z_i))'U_i] = O_p(n^{-1}) $ thus, $ \sqrt{n} E_n[h_n(Z_i)'U_i] $ satisfies the following Bahadur expansion:
\begin{equation}\label{eqn:Bahadur_Expansion}
	\sqrt{n} E_n[h_n(Z_i)'U_i] = \sqrt{n} E_n[h(Z_i)'U_i] + O_p(n^{-1/2}).
\end{equation}
\noindent From \eqref{eqn:Estimator_Expansion} and \eqref{eqn:Bahadur_Expansion}, the MMD is asymptotically linear thus rendering the multivariate Lindeberg-L\'evy Central Limit Theorem applicable. The covariance matrix estimator comprises the sample analogue of $ A $ and a consistent estimator of $B $. Denote $ \hat{B}_n \equiv  E_n[\hat{U}_i^2h_n(Z_i)'h_n(Z_i)] $ where $ \hat{U}_i \equiv Y_i - X_i\hat{\theta}_n $ is the MMD residual for observation $ i \in \{1,\ldots,n\} $. The covariance matrix estimator is given by $ \hat{A}_n^{-1}\hat{B}_n\hat{A}_n^{-1} $. The following theorem collects the asymptotic properties of the MMD viz. consistency, asymptotic normality, and consistency of the covariance matrix estimator.

\begin{theorem}\label{Theorem:Asymptotics}
	Under Assumptions \ref{Ass:DGP}-\ref{Ass:Identif}, (a) $ ||\hat{\theta}_n - \theta_o|| = O_p(n^{-1/2}) $; (b) $ \sqrt{n}(\hat{\theta}_n - \theta_o) \stackrel{d}{\rightarrow} \mathcal{N}(0, A^{-1}BA^{-1}) $; (c) $ \hat{A}_n^{-1}\hat{B}_n\hat{A}_n^{-1} \stackrel{p}{\rightarrow} A^{-1}BA^{-1} $.
\end{theorem}

\noindent Computationally, \Cref{Theorem:Asymptotics}(c) shows that the only departure of the MMD (and the linear ICM-class cast in the IV formulation \eqref{eqn:Estimator}) from the standard IV is the construction of the instrument $ \{h_n(Z_i), 1 \leq i \leq n\} $; estimation and inference are the same. Existing IV routines are, therefore, applicable without further modification.

 \subsection{Testing the Linear Completeness condition}\label{Sub_Sect:Relevance_Test}
The significantly weaker LC condition (\Cref{Ass:Identif}(b)) is helpful in addressing the problems of unavailable excluded instruments or potentially IV-weak but ICM-strong instruments. These advantages are only practically useful if the LC condition \Cref{Ass:Identif}(b) is testable. To this end, this paper generalises the \citet{escanciano2018simple} LC test to a multiple-covariate setting with a single endogenous covariate. This setting is quite prevalent in empirical practice.\footnote{The 2017 working paper version of  \citet{choi2021generalized} proposes a test of ICM relevance. The arguments of the authors rest on the structural model of interest and do not seem easily generalisable.} \footnote{For example, 101 out of 230 specifications (about 44\%) surveyed in \citet{andrews2019weak} (see page 732 of the paper) have a single endogenous covariate and a single excluded instrument. This means at least 101 of the specifications apply to this case since the number of excluded instruments in this case of the LC test is not limited to one; it can be zero, one, or more.} Partition $ X = [D, \tilde{X}] $ where $ \tilde{X} \in \mathbb{R}^{p_x-1} $ is a vector of exogenous covariates including the constant term and the univariate $ D $ is the endogenous covariate. Define $ \mathcal{E}^D(\eta) \equiv D - \tilde{X}\eta  $. The following result provides justification for casting the test of \Cref{Ass:Identif}(b) as an ICM specification test.
\begin{theorem}\label{Theorem:Rel_Test}
	Under \Cref{Ass:Bounds}(a), a test of \Cref{Ass:Identif}(b) in the presence of a single endogenous covariate $ D $ can be formulated using the hypotheses
	\begin{align*}
		\mathbb{H}_o &: \mathbb{P}( E[\mathcal{E}^D(\eta^*)|Z] = 0) = 1\text{ for some } \eta^* \in \mathbb{R}^{p_x-1};  \\
		\mathbb{H}_a &: \mathbb{P}( E[\mathcal{E}^D(\eta)|Z] = 0) < 1 \text{ for all } \eta \in \mathbb{R}^{p_x-1}. 
	\end{align*}
\end{theorem}
\noindent \Cref{Theorem:Rel_Test} in conjunction with Property (b) shows that the test of  \Cref{Ass:Identif}(b) in a single-endogenous-covariate setting simply involves a linear ICM regression of $ D $ on $ \tilde{X} $ with $ Z $ as instruments, and an ICM specification test. Rejecting $ \mathbb{H}_o $ at a nominal level provides evidence of non-parametric identification, i.e., \Cref{Ass:Identif}(b) holds. While the proposed LC test is simple and intuitive, it does not generalise to a multiple-endogenous-covariate setting in an obvious way. This is because the reasoning underpinning \Cref{Theorem:Rel_Test} only justifies setting $ D $ to at least one endogenous covariate without identifying any particular one. This pursuit is therefore left for future work.

\section{Monte Carlo Simulations}\label{Sect:Monte_Carlo_Sim}

Although ICM estimators have emerged since \citet{dominguez2004consistent}, no known study extensively compares their small sample performance. For instance, there is no small sample comparison of different ICM estimators in \citet{escanciano2006consistent}, \citet{antoine2014conditional}, or  \citet{escanciano2018simple}. In this regard, this section conducts extensive Monte Carlo simulations that examine the small sample performance of the MMD vis-\`a-vis existing ICM estimators. Following \citet{antoine2014conditional}, estimators of the K-class are also included. In addition, this section uses simulations to examine the small sample effect of the almost constant kernel problem on ICM estimators.\footnote{See the online supplement for further simulation results.} ICM estimators considered include the Weighted Minimum Distance (WMD) estimator and its modification \`a la \citet{fuller1977some} (WMDF) of \citet{antoine2014conditional}, the minimum distance estimator of \citet{dominguez2004consistent} (DL), the minimum distance estimator of \citet{escanciano2006consistent} (ESC6), and the Integrated Instrumental Variable (IIV) estimator of \citet{escanciano2018simple}. This section also considers the following estimators of the K-class: Two-Stage Least Squares (TSLS), the Jackknife Instrumental Variables Estimator (JIVE) of \citet{angrist1999jackknife}, the Limited Information Maximum Likelihood (LIML) estimator, the jackknife LIML (HLIM) of \citet{hausman2012instrumental}, and its modification \`a la \citet{fuller1977some} HFUL.

\subsection{Specifications}

Four different data-generating processes are considered. To examine the performance of estimators under the two features of ICM estimators, namely, IV regression without excluded instruments and IV regression with IV-weak but ICM-strong instruments, $ DGP_{0A} $, $ DGP_{0B} $, $ DGP_{1A} $, and $ DGP_{1B} $ are introduced. Define the function $ f_1(Z) \equiv \frac{2}{\sqrt{p_z}} \sum_{k=1}^{p_z}I\big(|Z_k| < - \Phi^{-1}(1/4)\big) $.

\begin{align*} DGP_{0A}:
	\begin{cases} 
		Y = \alpha_o + \beta_oD + \gamma_oZ + U\\
		D = 1/4+Z +\sqrt{\delta}Z^2 + V 
	\end{cases},\ & DGP_{0B}:
	\begin{cases} 
		Y = \alpha_o + \beta_oD_1 + \gamma_oD_2 + U\\
		D_1 = 1/4+Z +\sqrt{\delta}Z^2 + V/\sqrt{2}\\
		D_2=Z+ U/\sqrt{2} 
	\end{cases}
\end{align*}

\begin{align*} DGP_{1A}:
	\begin{cases} 
		Y = \alpha_o + \beta_oD + \gamma_o\tilde{X} + U\\
		D = 2\delta\Phi\big(\sum_{k=1}^{p_z}Z_k\big) + f_1(Z) + V 
	\end{cases},\ & DGP_{1B}:
	\begin{cases} 
		Y = \alpha_o + \beta_oD + \gamma_o\tilde{X} + U \\
		D = \sqrt{\delta} \frac{\mathrm{sin}(Z_1)\mathrm{sin}(Z_2)}{(1-\exp(-2))/4} + V 
	\end{cases}
\end{align*}

$ DGP_{0A} $ considers IV regression without excluded instruments with a single endogenous covariate, while $ DGP_{0B}$ extends $ DGP_{0A} $ to a two-endogenous-covariate setting. Estimators of the K-class are infeasible under $ DGP_{0A} $ and $ DGP_{0B} $ as they are not identified when $ p_z < p_x $. $ U $ and $ V $ are each standard normally distributed. $ \delta $ and $ \rho \equiv \mathrm{cov}[U,V] $ tune the strength of non-parametric identification and the degree of endogeneity respectively. For instance, $ \delta $ in $ DGP_{0A} $ tunes the non-linear mean dependence of $ D $ on $ Z $. In $ DGP_{1B} $, non-parametric identification is IV-irrelevant but ICM-relevant, and it depends on $ \delta $ since $ E[D|Z] = \sqrt{\delta} \frac{\mathrm{sin}(Z_1)\mathrm{sin}(Z_2)}{(1-\exp(-2))/4} \not = 0 $ a.s. for $ \delta \not = 0 $ but $ \mathrm{cov}[D,Z] = 0 $ for any $ \delta $. In $ DGP_{1A} $ (at $ \delta = 0 $) and $ DGP_{1B} $ (at $ \delta \not = 0 $), $ D $ is uncorrelated with but mean-dependent on $ Z $. For $ \delta \not = 0 $ in $ DGP_{1A} $, the linear projection of $ D $ on $ Z $ is largely dominated by its orthogonal complement; $ Z $ is thus IV-weak but ICM-strong in $ DGP_{1A} $ with $ \delta \in (0,1] $.\footnote{The first-stage $ F $-statistics are approximately $ 2.0 $ and $ 5.5 $ at $ \delta = 0.25 $ and $ \delta=0.5 $ respectively.}  $ \Phi(\cdot) $ denotes the cumulative distribution function of the standard normal distribution. In all DGPs, the set of instruments is generated as $ Z \sim \mathcal{N}(0,\Omega) $, where the  $ (k,l) $'th element of $ \Omega $ is $ \exp(-|k-l|) $. $ p_z = 2 $ in $ DGP_{1A} $ and $ DGP_{1B} $, while it is set to $ 1 $ in $ DGP_{0A} $, and $ DGP_{0B}$. $ \tilde{X} = Z_2 $ in both $ DGP_{1A} $ and $ DGP_{1B} $. $ \alpha_o=\beta_o=\gamma_o=1 $, and $ \rho = 0.5 $ in $ DGP_{0A}$ through $ DGP_{1B} $. 1000 Monte Carlo simulations are run for each specification.

\subsection{Bias and size control}\label{Sub_Sect:Bias_Control}
\Cref{Tab:Sim_0B_250,Tab:Sim_1B_250} present simulation results on $ DGP_{0A} $ through $ DGP_{1B} $ with sample size $ n=250 $.\footnote{Results for $ n=500 $ and other DGPs are available in the  online supplement.} In each table, $ \delta $ is varied in order to examine the performance of estimators at different degrees of non-parametric identification. For each level of $ \delta $, the mean bias (MB), median absolute deviation (MAD), the root mean square error (RMSE), and the empirical rejection rate (Rej.) for a 5\% $ t $-test of the null hypothesis  $ \mathbb{H}_o:\ \beta = \beta_o $ are reported.

\begin{table}[!htbp]
	\centering
	\setlength{\tabcolsep}{4pt}
	\caption{$ DGP_{0A} $ and $ DGP_{0B} $, $ n=250 $}
	\footnotesize
	\begin{tabular}{lcccccccccccc}
		\hline\hline
		& \multicolumn{4}{c}{$ \delta = 0.1 $} & \multicolumn{4}{c}{$ \delta = 0.5 $} & \multicolumn{4}{c}{$ \delta = 1.0 $} \\ 
		$ DGP_{0A} $ & MB &MAD & RMSE & Rej. & MB &MAD & RMSE & Rej. & MB &MAD & RMSE & Rej. \\
		\hline
		MMD & -0.024 & 0.098 & 0.163 & 0.044 & -0.005 & 0.044 & 0.067 & 0.058 & -0.003 & 0.030 & 0.047 & 0.060 \\
		WMD & -0.017 & 0.097 & 0.167 & 0.051 & -0.009 & 0.050 & 0.119 & 0.051 & 0.012 & 0.040 & 0.246 & 0.052 \\
		WMDF & -0.020 & 0.098 & 0.170 & 0.048 & -0.012 & 0.049 & 0.129 & 0.051 & 0.011 & 0.041 & 0.281 & 0.053 \\
		DL & -0.048 & 0.126 & 0.226 & 0.035 & -0.013 & 0.057 & 0.094 & 0.047 & -0.008 & 0.041 & 0.067 & 0.048 \\
		ESC6 & -0.038 & 0.112 & 0.190 & 0.030 & -0.009 & 0.047 & 0.074 & 0.040 & -0.005 & 0.033 & 0.051 & 0.042 \\
		IIV & -0.028 & 0.100 & 0.174 & 0.045 & -0.015 & 0.050 & 0.138 & 0.054 & -0.016 & 0.042 & 0.422 & 0.055 \\
		\hline
		
		$ DGP_{0B} $ &  &  &  &  &  &  &  &  &  &  &  &  \\ \cmidrule{1-1}
		
		MMD & 0.172 & 0.316 & 0.992 & 0.042 & 0.008 & 0.060 & 0.095 & 0.051 & 0.002 & 0.030 & 0.047 & 0.060 \\
		WMD & 1.765 & 0.301 & 51.31 & 0.043 & 0.010 & 0.062 & 0.126 & 0.052 & 0.004 & 0.039 & 0.104 & 0.056 \\
		WMDF & 0.187 & 0.308 & 1.312 & 0.043 & 0.011 & 0.062 & 0.130 & 0.050 & 0.005 & 0.039 & 0.101 & 0.056 \\
		DL & -1.353 & 0.380 & 52.332 & 0.043 & -0.006 & 0.079 & 0.126 & 0.047 & -0.007 & 0.041 & 0.066 & 0.046 \\
		ESC6 & 0.081 & 0.358 & 5.484 & 0.032 & 0.010 & 0.067 & 0.104 & 0.040 & 0.003 & 0.033 & 0.050 & 0.044 \\
		IIV & 0.246 & 0.319 & 2.162 & 0.039 & 0.017 & 0.061 & 0.173 & 0.051 & 0.007 & 0.040 & 0.096 & 0.056 \\
		\hline\hline
	\end{tabular}
	\footnotesize
	\label{Tab:Sim_0B_250}
\end{table}

\begin{table}[!htbp]
	\centering
	\setlength{\tabcolsep}{4pt}
	\caption{$ DGP_{1A} $ and $ DGP_{1B} $, $ n=250 $}
	\footnotesize
	\begin{tabular}{lcccccccccccc}
		\hline\hline
		& MB &MAD & RMSE & Rej. & MB &MAD & RMSE & Rej. & MB &MAD & RMSE & Rej. \\
		\hline
		$ DGP_{1A} $ & \multicolumn{4}{c}{$ \delta=0.0 $} & \multicolumn{4}{c}{$ \delta=0.25 $} & \multicolumn{4}{c}{$ \delta=0.5 $} \\ \cmidrule(l){1-1} \cmidrule(l){3-4} \cmidrule{7-8} \cmidrule(r){11-12}
		MMD & 0.000 & 0.058 & 0.088 & 0.045 & -0.001 & 0.058 & 0.088 & 0.047 & -0.001 & 0.057 & 0.087 & 0.047 \\
		WMD & -0.004 & 0.052 & 0.082 & 0.049 & -0.004 & 0.051 & 0.081 & 0.049 & -0.004 & 0.051 & 0.081 & 0.046 \\
		WMDF & -0.004 & 0.052 & 0.082 & 0.049 & -0.004 & 0.051 & 0.081 & 0.049 & -0.003 & 0.051 & 0.080 & 0.046 \\
		DL & 0.004 & 0.059 & 0.086 & 0.041 & 0.004 & 0.062 & 0.091 & 0.040 & 0.003 & 0.065 & 0.096 & 0.038 \\
		ESC6 & 0.002 & 0.049 & 0.078 & 0.045 & 0.001 & 0.049 & 0.078 & 0.048 & 0.001 & 0.048 & 0.076 & 0.049 \\
		IIV & -0.001 & 0.054 & 0.084 & 0.053 & -0.001 & 0.054 & 0.084 & 0.050 & -0.001 & 0.053 & 0.083 & 0.049 \\
		TSLS & 0.211 & 0.692 & 8.900 & 0.005 & -0.807 & 0.423 & 22.877 & 0.007 & -0.098 & 0.206 & 6.303 & 0.018 \\
		JIVE & -1.243 & 0.470 & 42.157 & 0.212 & -0.154 & 0.560 & 11.296 & 0.152 & -0.046 & 0.315 & 5.796 & 0.042 \\
		LIML & 0.211 & 0.692 & 8.900 & 0.005 & -0.807 & 0.423 & 22.877 & 0.007 & -0.098 & 0.206 & 6.303 & 0.018 \\
		HLIM & -1.243 & 0.470 & 42.157 & 0.212 & -0.154 & 0.560 & 11.296 & 0.152 & -0.046 & 0.315 & 5.796 & 0.042 \\
		HFUL & 0.416 & 0.853 & 43.279 & 0.010 & 24.514 & 0.495 & 785.699 & 0.008 & -0.088 & 0.219 & 3.054 & 0.013 \\

		\hline
		$ DGP_{1B} $ & \multicolumn{4}{c}{$ \delta=0.1 $} & \multicolumn{4}{c}{$ \delta=0.5 $} & \multicolumn{4}{c}{$ \delta=1.0 $} \\ \cmidrule(l){1-1} \cmidrule(l){3-4} \cmidrule{7-8} \cmidrule(r){11-12}
	
		MMD & 0.171 & 0.162 & 2.792 & 0.069 & 0.016 & 0.079 & 0.200 & 0.047 & 0.010 & 0.056 & 0.130 & 0.038 \\
		WMD & -0.006 & 0.079 & 0.118 & 0.031 & -0.003 & 0.034 & 0.049 & 0.037 & -0.002 & 0.024 & 0.035 & 0.036 \\
		WMDF & -0.003 & 0.078 & 0.117 & 0.033 & -0.002 & 0.034 & 0.049 & 0.039 & -0.002 & 0.023 & 0.035 & 0.037 \\
		DL & 0.073 & 0.177 & 3.362 & 0.068 & 0.026 & 0.091 & 0.475 & 0.038 & 0.065 & 0.065 & 1.508 & 0.035 \\
		ESC6 & 0.089 & 0.143 & 0.275 & 0.091 & 0.017 & 0.058 & 0.101 & 0.052 & 0.008 & 0.042 & 0.066 & 0.047 \\
		IIV & 0.032 & 0.092 & 0.127 & 0.059 & 0.005 & 0.039 & 0.056 & 0.049 & 0.001 & 0.028 & 0.040 & 0.048 \\
		TSLS & 0.376 & 0.733 & 9.171 & 0.005 & -0.065 & 0.483 & 14.444 & 0.000 & 0.209 & 0.363 & 5.691 & 0.000 \\
		JIVE & -0.619 & 0.400 & 35.945 & 0.302 & -0.413 & 0.187 & 13.922 & 0.187 & -0.101 & 0.123 & 5.889 & 0.128 \\
		LIML & 0.376 & 0.733 & 9.171 & 0.005 & -0.065 & 0.483 & 14.444 & 0.000 & 0.209 & 0.363 & 5.691 & 0.000 \\
		HLIM & -0.619 & 0.400 & 35.945 & 0.302 & -0.413 & 0.187 & 13.922 & 0.187 & -0.101 & 0.123 & 5.889 & 0.128 \\
		HFUL & -0.909 & 0.437 & 39.869 & 0.135 & 0.006 & 0.219 & 10.168 & 0.066 & -0.736 & 0.150 & 17.356 & 0.042 \\
		\hline\hline
	\end{tabular}
	\footnotesize
	\label{Tab:Sim_1B_250}
\end{table}

The results for $ DGP_{0A} $ and $ DGP_{0B} $ are presented in \Cref{Tab:Sim_0B_250}. In these DGPs, all estimators in the ICM-class perform well even though there is endogeneity without an excluded instrument. Notice particularly that the MMD compares favourably with other estimators within the ICM-class in terms of MAD and RMSE. For example, the MMD is not dominated in terms of RMSE at all levels of $ \delta $ and in terms of MAD at $ \delta \in \{0.5,1.0\} $ for $ DGP_{0A} $ and $ DGP_{0B} $. Size control is also generally good across all ICM estimators.

From \Cref{Tab:Sim_1B_250}, one notices in $ DGP_{1A} $ $ (\delta=0.0) $ and in $ DGP_{1B} $ that only estimators of the ICM-class have low bias and good size control as expected since $ Z $ is IV-irrelevant but ICM-relevant in these cases. The ESC6 and WMDF, in particular, are not dominated in terms of MAD and RMSE in $ DGP_{1A} $ and $ DGP_{1B} $, respectively. The ESC6 is, however, a little size-distorted at $ \delta=0.1 $ for $ DGP_{1B} $. Observe that even with $ \delta > 0 $ in $ DGP_{1A} $  where there is some linear dependence of $ D $ on $ Z $, the linear projection of $ D $ on $ Z $ is dominated by its  complement, whence the poor performance of the K-class of estimators. In spite of the poor performance of all estimators in the K-class, the JIVE and HLIM surprisingly show good size control at $ \delta=0.5 $ for $ DGP_{1A} $.

\subsection{Examining the almost constant kernel problem}

In this subsection, the almost constant kernel problem within the ICM-class of estimators comes into focus. To this end, $ DGP_{4} $ is specified below with $ \rho = \mathrm{cov}[U,V]=0.5 $ and $ p_z \in \{8, 18, 32\} $ at $ n \in \{250, 500, 1000\} $. Quite importantly, $ DGP_{4} $ allows for several relevant instruments where each instrument contributes a small fraction to the total strength of identification:

\begin{align*}
	DGP_{4}:
	\begin{cases} 
		Y = \alpha_o + \beta_oD + U\\
		D = \frac{1}{\sqrt{p_z}}\sum_{k=1}^{p_z}Z_k + V.
	\end{cases}
\end{align*}
\begin{table}[!htbp]
	\centering
	\setlength{\tabcolsep}{4pt}
	\caption{$ DGP_{4} $}
	\footnotesize
	\begin{tabular}{lcccccccccccc}
		\hline\hline
		& \multicolumn{4}{c}{$ p_z = 8 $} & \multicolumn{4}{c}{$ p_z = 18 $} & \multicolumn{4}{c}{$ p_z = 32 $} \\ 
		$ n=250 $ & MB &MAD & RMSE & Rej. & MB &MAD & RMSE & Rej. & MB &MAD & RMSE & Rej. \\ \cmidrule{1-1}
		\hline
		MMD & 0.007 & 0.034 & 0.048 & 0.061 & 0.014 & 0.031 & 0.047 & 0.072 & 0.023 & 0.036 & 0.051 & 0.126 \\
		WMD & -0.001 & 0.044 & 0.068 & 0.060 & -0.004 & 0.071 & 0.121 & 0.054 & 0.140 & 0.132 & 0.168 & 0.559 \\
		WMDF & 0.147 & 0.147 & 0.152 & 0.978 & 0.163 & 0.162 & 0.167 & 0.997 & 0.161 & 0.161 & 0.165 & 0.995 \\
		DL & 0.003 & 0.043 & 0.066 & 0.064 & 0.159 & 0.165 & 0.168 & 0.851 & 0.161 & 0.161 & 0.165 & 0.995 \\
		ESC6 & 0.007 & 0.034 & 0.048 & 0.058 & 0.013 & 0.031 & 0.047 & 0.069 & 0.022 & 0.034 & 0.050 & 0.115 \\
		IIV & 0.103 & 0.102 & 0.111 & 0.692 & 0.162 & 0.162 & 0.166 & 0.997 & 0.161 & 0.161 & 0.165 & 0.995 \\
		\hline

		$ n=500 $&  & &  &  &  & &  &  &  & &  &  \\ \cmidrule{1-1}

		MMD & 0.003 & 0.022 & 0.033 & 0.059 & 0.008 & 0.023 & 0.033 & 0.066 & 0.012 & 0.023 & 0.033 & 0.080 \\
		WMD & -0.002 & 0.031 & 0.046 & 0.066 & 0.002 & 0.048 & 0.072 & 0.049 & 0.136 & 0.131 & 0.153 & 0.753 \\
		WMDF & 0.103 & 0.103 & 0.107 & 0.946 & 0.164 & 0.164 & 0.166 & 1.000 & 0.161 & 0.161 & 0.163 & 1.000 \\
		DL & 0.001 & 0.030 & 0.045 & 0.055 & 0.125 & 0.131 & 0.140 & 0.640 & 0.161 & 0.161 & 0.163 & 1.000 \\
		ESC6 & 0.003 & 0.023 & 0.033 & 0.062 & 0.008 & 0.023 & 0.033 & 0.065 & 0.011 & 0.023 & 0.033 & 0.079 \\
		IIV & 0.071 & 0.071 & 0.079 & 0.592 & 0.163 & 0.163 & 0.165 & 1.000 & 0.161 & 0.161 & 0.163 & 1.000 \\
		\hline

		$ n=1000 $ &  &  &  &  &  &  &  &  &  &  &  &  \\ \cmidrule{1-1}

		MMD & 0.001 & 0.016 & 0.023 & 0.041 & 0.004 & 0.015 & 0.022 & 0.049 & 0.008 & 0.016 & 0.023 & 0.060 \\
		WMD & 0.000 & 0.018 & 0.026 & 0.042 & 0.002 & 0.024 & 0.034 & 0.046 & 0.123 & 0.120 & 0.129 & 0.906 \\
		WMDF & 0.040 & 0.039 & 0.045 & 0.396 & 0.164 & 0.163 & 0.165 & 1.000 & 0.161 & 0.161 & 0.162 & 1.000 \\
		DL & 0.000 & 0.020 & 0.030 & 0.042 & 0.064 & 0.066 & 0.087 & 0.294 & 0.162 & 0.162 & 0.163 & 1.000 \\
		ESC6 & 0.001 & 0.016 & 0.023 & 0.042 & 0.004 & 0.015 & 0.022 & 0.047 & 0.007 & 0.016 & 0.023 & 0.056 \\
		IIV & 0.035 & 0.035 & 0.042 & 0.295 & 0.162 & 0.161 & 0.163 & 1.000 & 0.161 & 0.161 & 0.162 & 1.000 \\
		\hline\hline
	\end{tabular}
	\footnotesize
	\label{Tab:Sim_4_1000}
\end{table}

For all $ p_z $ considered, notice from \Cref{Tab:Sim_4_1000} that MAD and RMSE are non-increasing in $ n $ for all estimators. The MMD and ESC6, specifically, have MAD and RMSE that are strictly decreasing in sample size. The performance of the MMD and ESC6 in terms of MAD and RMSE are quite indistinguishable. Only the MMD and ESC6 show meaningful improvement with sample size at $ p_z=32 $. For example, the MAD of the WMDF, DL, and IIV at $ p_z = 32 $ is practically unchanged across sample sizes, and doubling the sample size from $ 500 $ to $ 1000 $ leaves the RMSE of the DL at $ p_z=32 $ practically unchanged. At each sample size, the MAD and RMSE of the MMD and ESC6 are very stable across $ p_z $. In fact, $ \sqrt{n}\times $MAD for the MMD and ESC6 are approximately $ 0.5 $ across all $ n $ and $ p_z $ which is indicative of $ \sqrt{n} $-consistency. This is, however, not the case for the WMD, WMDF, DL, and IIV as $ \sqrt{n}\times $MAD is sensitive to $ p_z $ at all $ n $ and sensitive to $ n $ at $ p_z\in \{18,32\} $. As a case in point, $ \sqrt{n}\times $MAD of the WMDF, and IIV are increasing in $ n $ at $ p_z\in \{18, 32\} $. It, however, ought to be emphasised that these are small sample issues as this paper considers $ p_z $ fixed with respect to $ n $.

At $ (n, p_z) =(250, 32) $, all estimators suffer substantial size distortion, although that of the MMD and ESC6 is much less severe. The size distortion of the MMD and ESC6 declines with the sample size, whereas that of the other estimators (which have multiplicative kernels) worsens with sample size. In fact, the empirical sizes of the WMDF and IIV at ($ n\in \{500,1000\}, \ p_z = 18 $) and the WMDF, DL, and IIV at ($ n\in \{500,1000\}, \ p_z=32 $) equal 1. Even at $ p_z=8 $, the size distortion of the WMDF and IIV remains severe across sample sizes. The size distortion of the WMD is not as pronounced as that of the other ICM estimators with multiplicative kernels. This is perhaps attributable to the jackknifed LIML-like structure of its kernel which is intended to mitigate dispersion -- see \citet[eqn. 14]{antoine2014conditional}. This small simulation exercise shows the adverse effect of the almost constant kernel problem on the quality of inference in the ICM-class of estimators with multiplicative kernels. The almost constant kernel problem appears to increase finite sample bias and induce severe size distortions in ICM estimators with multiplicative kernels. In comparison to results in \Cref{Sub_Sect:Bias_Control}, one observes that multiplicative-kernel ICM estimators, namely, the WMD, DL, and IIV, perform reasonably well when the dimension of $ Z $ is small, say one or two. Their relative performance deteriorates for moderate dimensions of $ Z $.

\section{Empirical Example}\label{Sect:Empirical_Example}

This section presents an empirical example from \citet{hornung2014immigration} which illustrates the practical usefulness of the MMD estimator. Quite importantly, the endogenous covariate is non-linearly mean-dependent on exogenous instruments and hence lends itself to IV regression without an excluded instrument using an ICM estimator. The motivation for using the MMD in an empirical application where an excluded instrument is available is to verify the reliability of the MMD without an excluded instrument in a real-data setting. Also, the sample size is small $ (n=150) $, the instruments are not very IV-strong, and the instrument set is moderately sized $ (p_z = 11) $.

\citet{hornung2014immigration} seeks to identify the long-term impact of Huguenot skilled-worker migration on the  productivity of textile manufactories in Prussia. The outcome variable is the value of a firm's output in a given town, and the covariate of interest is the population share of Huguenots in a town (Percent Huguenots). Other covariates include the number of workers, the number of looms, the value of material input, and regional and town-specific characteristics that can impact productivity -- see \citet{hornung2014immigration} for details. As Huguenots are highly specialised in the textile industry, one can expect the population share of Huguenots (the endogenous covariate) to, for example,  depend on the number of looms. As one cannot, \emph{a priori}, rule out non-linearities in the dependence structure of covariates, IV without excludability is possible. Excluded instruments proposed by \citet{hornung2014immigration} for \emph{Percent Huguenots} include population losses during the Thirty Years' War and its interpolated version -- see \citet{hornung2014immigration} for details.

\begin{table}[!htbp]
	\setlength{\tabcolsep}{4pt}
	\caption{Estimates - Huguenot Productivity -  \citet{hornung2014immigration} }
	\begin{center}
	\begin{tabular}{lccccccc}
		\hline\hline
		& MMD & OLS & MMD & IV & MMD & IV & MMD* \\ 
		& (1) & (1) & (2) & (2) & (3) & (3) & (4) \\
		\hline\hline
		Percent Huguenots &1.823 & 1.741 & 1.935 & 3.475 & 1.928 & 3.38 & 1.816 \\ 
		& (0.797) & (0.471) & (0.888) & (0.960) & (0.869) & (0.990) & (0.806) \\
		Excluded Instr. &  &  & $ \checkmark $ & $ \checkmark $ & $ \checkmark $ & $ \checkmark $ & \\\hline
		$ \underline{\text{Relevance}} $ &  &  &  & & & & \\
		\multicolumn{1}{r}{\textit{LC pval}} &  &  & \textit{0.001} &  & \textit{0.002} &  & \textit{0.084} \\
		\multicolumn{1}{r}{\textit{KP F-Stat}} &  &  &  & \textit{14.331} & & \textit{22.531} & \\
		\hline
		$ \underline{\text{Specification}} $ &  &  &  & & & & \\
		\multicolumn{1}{r}{\textit{S\&Z pval}} & 0.300 & 0.490 & 0.327 & 0.296 & 0.317 & 0.340 & 0.300 \\
		\hline\hline
	\end{tabular}
\end{center}
	\footnotesize
	\renewcommand{\baselineskip}{11pt}
	\textbf{Note:} The number of observations in each specification is 150 firms. Heteroskedasticity-robust standard errors are given in parentheses. 999 wild-bootstrap samples are used to conduct the LC and \citet{su2017martingale} (\textit{S\&Z}) specification tests. Excluded Instr. denotes whether an excluded instrument is used, and the KP $ F $-Stat. denotes \citeauthor{kleibergen2006generalized}'s \citeyearpar{kleibergen2006generalized} rk $ F $-statistic. The last column MMD* implements the MMD without the excluded instrument. The data are sourced from \citet{hornung2014data}.
	\label{Tab:Hornung2014_Hetero}
\end{table}

\Cref{Tab:Hornung2014_Hetero} presents the empirical results. The upper panel presents coefficient estimates with standard errors, the second presents p-values of the proposed LC test, the $ F $-statistic of \citet{kleibergen2006generalized} rank test, and the third presents the specification test of \citet{su2017martingale}. There are four specifications in \Cref{Tab:Hornung2014_Hetero}. The first compares the MMD to the OLS, the second uses population losses during the Thirty Years' War as the excluded instrument, the third uses its interpolated version, and the fourth runs MMD without instrumenting for the endogenous covariate.\footnote{MMD*(4) thus differs from MMD(1) by the exclusion (without replacement) of the endogenous covariate from the set of instruments.}

One observes that MMD estimates and standard errors are stable across specifications, while OLS/IV estimates show substantial variation. In the presence of excluded instruments (specifications (2) and (3)), MMD estimates are more precise than the IV. Interestingly, in the last column, where no excluded instrument is used, the MMD estimate is reasonably close to MMD estimates with excluded instruments and appears to be more precisely estimated than the IV and the MMD itself with excluded instruments. This shows that the unavailability of excluded instruments in this empirical example hurts neither the plausibility of the estimate nor its precision in a meaningful way.

\section{Conclusion}\label{Sect:Conclusion}

This paper introduces a linear IV estimator to the ICM-class of estimators that minimises the mean dependence of an error term on a finite set of instruments. The proposed estimator, like that of \citet{escanciano2006consistent}, is more robust to the almost constant kernel problem relative to estimators within the ICM-class with multiplicative kernels. This paper highlights a remarkable feature of the ICM-class that can address the challenge of unavailable excluded instruments. Given that the practical usefulness of ICM estimators in tackling the aforementioned challenge crucially lies in the testability of the LC condition, this paper shows that the LC test for ICM estimators can be cast as a standard ICM specification test. The estimator's closed-form expression makes it computationally fast to implement using available linear IV routines.

The type of estimator proposed in this paper enables the researcher to exploit unknown forms of identifying variation that cannot be exploited using conventional IV methods. Although this approach is not automatically applicable whenever excluded instruments are unavailable, nor is it a panacea for all forms of the weak IV problem, the testability of the LC condition makes the approach practically useful as a researcher is able to ascertain applicability in a given empirical context. Firstly, the empirical example in this paper shows an empirically relevant scenario where an ICM estimator still yields reliable estimates when the researcher lacks excluded instruments but likely faces an endogeneity problem. Secondly, the empirical example helps to demonstrate the robustness and reliability of inference that the MMD and the ICM-class as a whole offers. The simulation exercise offers insights into how ICM estimators can rescue a project from the otherwise hard-to-solve problem of unavailable excluded instruments or very IV-weak (but ICM-strong) instruments. It shows a favourably competitive performance of the proposed MMD relative to other estimators of the ICM- and K-classes, and calls attention to the severity of the almost constant kernel problem in multiplicative-kernel ICM-estimators. Although not entirely new to the literature, the characterisation of the almost constant kernel problem in this paper is, nonetheless, interesting as it sheds light on the case of a moderately large but fixed dimension of $ Z $.

This paper leaves a number of interesting avenues for future work, e.g., extending the current framework to non-linear models, models with non-smooth objective functions such as quantile regression, and multivariate outcomes. It will also be interesting to extend the LC test to a multiple-endogenous-covariate setting. Owing to the frequency of autocorrelation and clustering in empirical work, it will be interesting to extend the current $ iid $ framework for autocorrelation- and cluster-robust inference.

\section*{Acknowledgements}
An earlier draft of this paper circulated under the title \textit{IV Regression with Possibly Uncorrelated Instruments}. The author acknowledges very useful feedback from the Editor and two anonymous referees. This paper also benefited from the invaluable comments of Al-mouksit Akim, Firmin Ayivodji, Brantly Callaway, Weige Huang, Feiyu Jiang, Gilles Koumou, Abdul-Nasah Soale, Guy Tchuente, and participants at the 2021 Latin American Meeting of the Econometric Society (LAMES), 2022 North American Summer Meeting of the Econometric Society (NASMES), and the 2021/2022 Université Mohammed VI Polytechnique AIRESS Seminar Series.\\


    \section*{Appendix: Proof of Theorem 3.1}
    \renewcommand{\theequation}{A.\arabic{equation}}
    \renewcommand{\thesection}{A}
    \setcounter{equation}{0}

		\medskip
		
	The proof of \Cref{Theorem:Asymptotics} draws on Lemmata in the online supplement.
		
	\textbf{Part (a)}: By the continuity of the inverse at a non-singular matrix, \Cref{Ass:Identif}(b), Lemma S.2, and the continuous mapping theorem, $ \hat{A}_n^{-1} = A^{-1} + o_p(1) $. Thus, substituting $ Y_i $ from the data-generating process (\Cref{Ass:DGP}) into the estimator \eqref{eqn:Estimator} and that $ E_n[(h_n(Z_i)-h(Z_i))'U_i] = O_p(n^{-1}) $ by Lemma S.3,
	
	\begin{align}\label{eqn:theta_bias}
		\hat{\theta}_n - \theta_o  &= \hat{A}_n^{-1} E_n[h_n(Z_i)'U_i] \nonumber \\
		&= A^{-1} E_n[h(Z_i)'U_i] + O_p(n^{-1})\times A^{-1} + o_p(1)\times  E_n[h(Z_i)'U_i] + o_p(n^{-1}) \nonumber\\
		&= A^{-1} E_n[h(Z_i)'U_i] + o_p(n^{-1/2}).
	\end{align}
	
	\noindent $|| \mathrm{var}[  E_n[h(Z_i)'U_i] ]|| = || \mathrm{var}[\sqrt{n} E_n[h(Z_i)'U_i] ]||/n = ||B||/n \leq 4\bar{\sigma}^2C/n  $ under the assumptions of Lemma S.1 implies $ || E_n[h(Z_i)'U_i]|| = O_p(n^{-1/2}) $ by the Markov inequality. From \eqref{eqn:theta_bias}, $ ||\hat{\theta}_n - \theta_o|| = O_p(n^{-1/2}) $.
	\hfill$\square$\\
	
	
	\textbf{Part (b)}: \eqref{eqn:theta_bias} shows that the MMD is asymptotically linear, thus
	\begin{equation}\label{eqn:root_n_consistency}
		\begin{split}
			\sqrt{n}(\hat{\theta}_n - \theta_o)  &= A^{-1}\sqrt{n} E_n[h(Z_i)'U_i] + o_p(1).
		\end{split}
	\end{equation}
	
	\noindent As $  E[U^2||h(Z)||^2] =  E[\sigma^2(Z)||h(Z)||^2] $ by the LIE, it follows from (S.3)  that $  E[U^2||h(Z)||^2] \leq 4\bar{\sigma}^2C $. In addition to Assumptions \ref{Ass:DGP} and \ref{Ass:Identif}(a), $ \sqrt{n} E_n[h(Z_i)'U_i]  \xrightarrow{d} \mathcal{N}(0,B) $ by the multivariate Lindeberg-L\'evy Central Limit Theorem. The conclusion follows from the linearity of $ \sqrt{n}(\hat{\theta}_n - \theta_o) $ in $ \sqrt{n} E_n[h(Z_i)'U_i] $ in  \eqref{eqn:root_n_consistency}.
	\hfill$\square$\\
	
	
	\textbf{Part (c)}: First $\displaystyle \plim_{n \rightarrow \infty} \hat{A}_n=A $ by Lemma S.2. Second, define $B_n \equiv  E_n[U_i^2h(Z_i)'h(Z_i)] $. $  E[B_n]=B $, and by the triangle inequality, 
	\begin{equation}\label{eqn:Decomp_Bhat}
		||\hat{B}_n-B|| \leq ||\hat{B}_n-B_n|| + ||B_n-B||.
	\end{equation}	From the Schwartz inequality and the proof of \Cref{Theorem:Asymptotics}(b) above, $  E[||U^2h(Z)'h(Z)||] \leq  E[U^2||h(Z)||^2] \leq 4\bar{\sigma}^2C, $
	hence $ ||B_n - B|| = o_p(1) $ by the Weak Law of Large Numbers (WLLN). From \eqref{eqn:Decomp_Bhat}, it suffices to show that $ ||\hat{B}_n-B_n|| = o_p(1) $. 
	
	Consider the decomposition 
	\begin{align*}
		\hat{B}_n-B_n &=  E_n[\hat{U}_i^2(h_n(Z_i)-h(Z_i))'h_n(Z_i)] +  E_n[\hat{U}_i^2h(Z_i)'(h_n(Z_i)-h(Z_i))]\\ 
		&+  E_n[(\hat{U}_i^2-U_i^2)h(Z_i)'h(Z_i)].
	\end{align*}
	 Apply the triangle inequality twice and the Schwartz inequality to obtain  
	\begin{equation}\label{eqn:Bhat_B_Bound1}
		\begin{split}
			||\hat{B}_n-B_n|| 
			&\leq  E_n[\hat{U}_i^2||h_n(Z_j)-h(Z_j)|| \cdot(||h_n(Z_i)||+||h(Z_i)||)] +  E_n[|\hat{U}_i^2-U_i^2|\cdot||h(Z_i)||^2].
		\end{split}
	\end{equation} It follows from \eqref{eqn:Bhat_B_Bound1} and the equality $ \hat{U}_i^2-U_i^2 = (\hat{U}_i-U_i)^2 + 2U_i(\hat{U}_i-U_i)$ that
	\begin{equation}\label{eqn:Bhat_B_Bound2}
		\begin{split}
			||\hat{B}_n-B_n|| &\leq  E_n[\hat{U}_i^2||h_n(Z_j)-h(Z_j)|| \cdot(||h_n(Z_i)||+||h(Z_i)||)]\\
			&+ \max_{1\leq j \leq n}|\hat{U}_j-U_j|^2\cdot E_n[||h(Z_i)||^2] + 2\max_{1\leq j \leq n}|\hat{U}_j-U_j|\cdot E_n[|U_i|\cdot||h(Z_i)||^2].
		\end{split}
	\end{equation} 
	
	From Lemma S.5, each summand of \eqref{eqn:Bhat_B_Bound2} is $ o_p(1) $, and the proof is complete.
	\hfill$\square$\\
	
	\vspace{1cm}
	
		\textit{The online appendix is available on the author's website.} 	


\begin{thebibliography}{}
	
	\bibitem[\protect\citeauthoryear{Andrews et al.}{Andrews et al.}{2019}]{andrews2019weak}
	Andrews, I., J. H.~Stock and L.~Sun (2019).
	\newblock Weak instruments in instrumental variables regression: Theory and Practice.
	\newblock {\em Annual Review of Economics\/} {\em 11}, 727--753.
	
	\bibitem[\protect\citeauthoryear{Angrist et al.}{Angrist et al.}{1999}]{angrist1999jackknife}
	Angrist, J. D., G. W.~Imbens and A. B.~Krueger (1999).
	\newblock Jackknife instrumental variables estimation.
	\newblock {\em Journal of Applied Econometrics\/} {\em 14}, 57--67.
	
	\bibitem[\protect\citeauthoryear{Antoine and Lavergne}{Antoine and Lavergne}{2014}]{antoine2014conditional}
	Antoine, B.  and P.~Lavergne (2014).
	\newblock Conditional moment models under semi-strong identification.
	\newblock {\em Journal of Econometrics\/} {\em 182}, 59--69.
		
	\bibitem[\protect\citeauthoryear{Antoine and Sun}{Antoine and Sun}{2022}]{antoine2022partially}
	Antoine, B. and X.~Sun (2022).
	\newblock Partially linear models with endogeneity: a conditional moment-based approach.
	\newblock {\em Econometrics Journal\/} {\em 25}, 256--275.
		
	\bibitem[\protect\citeauthoryear{Bierens}{Bierens}{1982}]{bierens1982consistent}
	Bierens, H.~J.  (1982).
	\newblock Consistent model specification tests.
	\newblock {\em Journal of Econometrics\/} {\em 20}, 105--134.
	
	\bibitem[\protect\citeauthoryear{Bierens}{Bierens}{1990}]{bierens1990consistent}
	Bierens, H.~J.  (1990).
	\newblock A consistent conditional moment test of functional form.
	\newblock  {\em Econometrica\/} {\em 58}, 1443--1458.
	
	\bibitem[\protect\citeauthoryear{Bierens and Ploberger}{Bierens and Ploberger}{1997}]{bierens1997asymptotic}
	Bierens, H.~J. and W.~Ploberger (1997).
	\newblock Asymptotic theory of integrated conditional moment tests.
	\newblock {\em Econometrica\/} {\em 65}, 1129--1151.
	
	\bibitem[\protect\citeauthoryear{Canay et al.}{Canay et al.}{2013}]{canay2013testability}
	Canay, I.~A., A. Santos, and A.~M. Shaikh (2013).
	\newblock On the testability of identification in some nonparametric models with endogeneity.
	\newblock {\em Econometrica\/} {\em 81}, 2535--2559.
	
	\bibitem[\protect\citeauthoryear{Carrasco and Tchuente}{Carrasco and Tchuente}{2015}]{carrasco2015regularized}
	Carrasco, M. and G. Tchuente (2015).
	\newblock Regularized LIML for many instruments.
	\newblock {\em Journal of Econometrics\/} {\em 186}, 427--442.
	
	\bibitem[\protect\citeauthoryear{Choi et al.}{Choi et al.}{2022}]{choi2021generalized}
	Choi, J., J.~C. Escanciano, and J. Guo (2022).
	\newblock Generalized band spectrum estimation with an application to the new Keynesian Phillips Curve.
	\newblock {\em Journal of Applied Econometrics\/} {\em 37}, 1055-1078.
		
	\bibitem[\protect\citeauthoryear{Dom\'i­nguez and Lobato}{Dom\'i­nguez and Lobato}{2004}]{dominguez2004consistent}
	Dom\'inguez, M.~A.  and I.~N. Lobato (2004).
	\newblock Consistent estimation of models defined by conditional moment restrictions.
	\newblock {\em Econometrica} {\em 72}, 1601--1615.
	
	\bibitem[\protect\citeauthoryear{Dom\'i­nguez and Lobato}{Dom\'i­nguez and Lobato}{2015}]{dominguez2015simple}
	Dom\'inguez, M.~A.  and I.~N. Lobato (2015).
	\newblock A simple omnibus overidentification specification test for time series econometric models.
	\newblock {\em Econometric Theory} {\em 31}, 891--910.
		
	\bibitem[\protect\citeauthoryear{Donald and Newey}{Donald and Newey}{2001}]{donald2001choosing}
	Donald, S.~G. and W.~K. Newey (2001).
	\newblock Choosing the number of instruments.
	\newblock {\em Econometrica\/} {\em 69}, 1161--1191.
	
	\bibitem[\protect\citeauthoryear{Escanciano}{Escanciano}{2006}]{escanciano2006consistent}
	Escanciano, J.~C. (2006).
	\newblock A consistent diagnostic test for regression models using projections.
	\newblock {\em Econometric Theory\/} {\em 22}, 1030--1051.
	
	\bibitem[\protect\citeauthoryear{Escanciano}{Escanciano}{2018}]{escanciano2018simple}
	Escanciano, J.~C. (2018).
	\newblock A simple and robust estimator for linear regression models with strictly exogenous instruments.
	\newblock {\em Econometrics Journal\/} {\em 21}, 36--54.
	
	\bibitem[\protect\citeauthoryear{Escanciano and Terschuur}{Escanciano and Terschuur}{2022}]{escanciano2022debiased}
	Escanciano, J.~C. and J.~R. Terschuur (2022).
	\newblock Debiased semiparametric U-statistics: Machine learning inference on inequality of opportunity.
	\newblock Working paper, arXiv preprint arXiv:2206.05235.
	
	\bibitem[\protect\citeauthoryear{Fuller}{Fuller}{1977}]{fuller1977some}
	Fuller, W.~A. (1977).
	\newblock Some properties of a modification of the limited information estimator.
	\newblock {\em Econometrica\/} {\em 45}, 939--953.
	
	\bibitem[\protect\citeauthoryear{Jiang and Tsyawo}{Jiang and Tsyawo}{2022}]{jiangTsyawo2022aconsistent}
	Jiang, F. and E.~S. Tsyawo (2022).
	\newblock A Consistent ICM-based $\chi^2$ Specification Test.
	\newblock Working paper, arXiv preprint arXiv:2208.13370
		
	\bibitem[\protect\citeauthoryear{Hansen}{Hansen}{2021}]{hansen_2021econometrics}
	Hansen, B.~E.  (2021).
	\newblock {\em Econometrics\/}.
	\newblock University of Wisconsin, Dept. of Economics.
	
	\bibitem[\protect\citeauthoryear{Hansen and Kozbur}{Hansen and Kozbur}{2014}]{hansen2014instrumental}
	Hansen, C. and D. Kozbur (2014).
	\newblock Instrumental variables estimation with many weak instruments using
	regularized JIVE.
	\newblock {\em Journal of Econometrics\/} {\em 182}, 290--308.
	
	\bibitem[\protect\citeauthoryear{Hausman et al.}{Hausman et al.}{2012}]{hausman2012instrumental}
	Hausman, J.~A., W.~K. Newey, T. Woutersen, J.~C. Chao, and N.~R. Swanson (2012).
	\newblock Instrumental variable estimation with heteroskedasticity and many
	instruments.
	\newblock {\em Quantitative Economics\/} {\em 3}, 211--255.
	
	\bibitem[\protect\citeauthoryear{Hornung}{Hornung}{2014a}]{hornung2014immigration}
	Hornung, E. (2014a).
	\newblock Immigration and the diffusion of technology: The Huguenot diaspora in Prussia.
	\newblock {\em American Economic Review} {\em 104}, 84--122.
	
	\bibitem[\protect\citeauthoryear{Hornung}{Hornung}{2014b}]{hornung2014data}
	Hornung, E. (2014b).
	\newblock Replication data for: Immigration and the diffusion of technology: The Huguenot diaspora in Prussia.
	\newblock Data retrieved from OPENICPSR, 2019-10-11, https://doi.org/10.3886/E112731V1.
	
	\bibitem[\protect\citeauthoryear{Kim et al.}{Kim et al.}{2020}]{kim2020robust}
	Kim, I., S. Balakrishnan, and L. Wasserman (2020).
	\newblock Robust multivariate nonparametric tests via projection averaging.
	\newblock {Annals of Statistics\/} {\em 48}, 3417--3441.
	
	\bibitem[\protect\citeauthoryear{Kleibergen and Paap}{Kleibergen and Paap}{2006}]{kleibergen2006generalized}
	Kleibergen, F. and R. Paap (2006).
	\newblock Generalized reduced rank tests using the singular value decomposition.
	\newblock {\em Journal of Econometrics\/} {\em 133}, 97--126.
	
	\bibitem[\protect\citeauthoryear{Lee}{Lee}{1990}]{lee1990u}
	Lee, A.~J. (1990).
	\newblock {\em U-statistics: Theory and Practice\/}.
	\newblock Boca Raton, FL: CRC Press.
	
	\bibitem[\protect\citeauthoryear{Newey and Powell}{Newey and Powell}{2003}]{newey2003instrumental}
	Newey, W.~K.  and J.~L. Powell (2003).
	\newblock Instrumental variable estimation of nonparametric models.
	\newblock {\em Econometrica\/} {\em 71}, 1565--1578.
		
	\bibitem[\protect\citeauthoryear{Shao and Zhang}{Shao and Zhang}{2014}]{shao2014martingale}
	Shao, X. and J. Zhang (2014).
	\newblock Martingale difference correlation and its use in high-dimensional variable screening.
	\newblock {\em Journal of the American Statistical Association} {\em 109}, 1302--1318.
	
	\bibitem[\protect\citeauthoryear{Shin}{Shin}{2008}]{shin2008semiparametric}
	Shin, Y. (2008).
	\newblock Semiparametric estimation of the box--cox transformation model.
	\newblock {\em Econometrics Journal\/} {\em 11}, 517--537.
	
	\bibitem[\protect\citeauthoryear{Stock and Yogo}{Stock and Yogo}{2005}]{stock2005testing}
	Stock, J. and M. Yogo (2005).
	\newblock Testing for Weak Instruments in Linear IV Regression
	\newblock In Andrews DWK (Ed.) {\em Identification and Inference for Econometric Models},
	80--108. New York, NY: Cambridge University Press.
	
	\bibitem[\protect\citeauthoryear{Stute}{Stute}{1997}]{stute1997nonparametric}
	Stute, W. (1997).
	\newblock Nonparametric model checks for regression.
	\newblock {\em Annals of Statistics\/} {\em 25}, 613--641.
	
	\bibitem[\protect\citeauthoryear{Su and Zheng}{Su and Zheng}{2017}]{su2017martingale}
	Su, L. and X. Zheng (2017).
	\newblock A martingale-difference-divergence-based test for specification.
	\newblock {\em Economics Letters} {\em 156}, 162--167.
	
	\bibitem[\protect\citeauthoryear{Sun}{Sun}{2022}]{sun2022estimation}
	Sun, X. (2022).
	\newblock Estimation of Heterogeneous Treatment Effects Using a Conditional Moment Based Approach.
	\newblock Working paper, arXiv preprint arXiv:2210.15829.
	
	\bibitem[\protect\citeauthoryear{Sz\'ekely and Rizzo}{Sz\'ekely and Rizzo}{2012}]{szekely2012uniqueness}
	Sz\'ekely, G.~J.  and M.~L. Rizzo (2012).
	\newblock On the uniqueness of distance covariance.
	\newblock {\em Statistics and Probability Letters\/} {\em 82}, 2278--2282.
	
	\bibitem[\protect\citeauthoryear{Sz\'ekely et al.}{Sz\'ekely et al.}{2007}]{szekely2007measuring}
	Sz\'ekely, G.~J., M.~L. Rizzo and N.~K. Bakirov (2007).
	\newblock Measuring and testing dependence by correlation of distances.
	\newblock {\em Annals of Statistics\/} {\em 35}, 2769--2794.
	
	\bibitem[\protect\citeauthoryear{Wang}{Wang}{2018}]{wang2018consistent}
	Wang, X. (2018).
	\newblock Consistent Estimation Of Models Defined By Conditional Moment Restrictions Under Minimal Identifying Conditions.
	\newblock Working Paper, Xiamen University.
	
	\bibitem[\protect\citeauthoryear{Wooldridge}{Wooldridge}{2010}]{wooldridge2010econometric}
	Wooldridge, J.~M. (2010).
	\newblock {\em Econometric analysis of cross section and panel data\/} (2nd ed.).
	\newblock Cambridge, MA: MIT Press.
		
\end{thebibliography}
\end{document}